\documentclass[aps,pra,reprint,superscriptaddress,showpacs]{revtex4-1}
\usepackage[utf8]{inputenc}
\usepackage{graphicx}

\begin{document}


\title{Photoionization of metastable heliumlike C$^{4+}(1s 2s~^3S_1)$ ions:\\ Precision study
of intermediate doubly excited states}

\author{A.~M\"{u}ller}
\email[]{Alfred.Mueller@iamp.physik.uni-giessen.de}
\affiliation{Institut f\"{u}r Atom- und Molek\"{u}lphysik, Justus-Liebig-Universit\"{a}t Gie{\ss}en, Leihgesterner Weg 217, 35392 Giessen, Germany}
\author{E. Lindroth}
\affiliation{Department of Physics, Stockholm University, Alba Nova University Center, 
106 91 Stockholm, Sweden}
\author{S.~Bari}
\affiliation{FS-SCS, DESY, Notkestr. 85,  22607  Hamburg, Germany}
\author{A.~Borovik Jr.}
\affiliation{I. Physikalisches Institut, Justus-Liebig-Universit\"{a}t Gie{\ss}en, Leihgesterner Weg 217, 35392  Giessen, Germany}
\author{P.-M.~Hillenbrand}
\altaffiliation{present address: Columbia Astrophysics Laboratory, Columbia University, 550 West 120th Street, New York, NY 10027, USA}
\affiliation{GSI Helmholtzzentrum f\"ur Schwerionenforschung, Planckstr. 1, 64291 Darmstadt, Germany}
\author{K.~Holste}
\affiliation{I. Physikalisches Institut, Justus-Liebig-Universit\"{a}t Gie{\ss}en, Heinrich-Buff-Ring 16, 35392  Giessen, Germany}
\author{P. Indelicato}
\affiliation{Laboratoire Kastler Brossel, Sorbonne Universit\'{e},
CNRS, ENS-PSL Research University, Coll\`{e}ge de France,  Case\ 74;\ 4, place Jussieu, 75005 Paris, France}
\author{A.~L.~D.~Kilcoyne}
\affiliation{Advanced Light Source, Lawrence Berkeley National Laboratory, 1 Cyclotron Road, M.S. 7R0222, Berkeley, CA 94720-8229, USA}
\author{S.~Klumpp}
\affiliation{DESY Photon Science, FS-FLASH-D, Notkestr. 85, 22607  Hamburg, Germany}
\author{M.~Martins}
\affiliation{Institut f\"{u}r Experimentalphysik, Universit\"{a}t Hamburg, Luruper Chaussee 149, 22761  Hamburg, Germany}
\author{J.~Viefhaus}
\altaffiliation{Present address: Helmholtz-Zentrum Berlin, Department Optics and Beamlines, BESSY II, Albert-Einstein-Str. 15,  12489  Berlin, Germany}
\affiliation{DESY Photon Science, FS-PE, Notkestr. 85, 22607  Hamburg, Germany}
\author{P.~Wilhelm}
\altaffiliation{present address: Max-Planck-Institut f\"{u}r Kernphysik, Saupfercheckweg 1, 69029 Heidelberg, Germany}
\affiliation{Institut f\"{u}r Atom- und Molek\"{u}lphysik, Justus-Liebig-Universit\"{a}t Gie{\ss}en, Leihgesterner Weg 217, 35392 Giessen, Germany}
\author{ S.~Schippers}
\affiliation{I. Physikalisches Institut, Justus-Liebig-Universit\"{a}t Gie{\ss}en, Leihgesterner Weg 217, 35392  Giessen, Germany}

\date{\today}

\begin{abstract}
In a joint experimental and theoretical endeavour, photoionization of metastable C$^{4+}(1s 2s~^3S_1)$ ions via intermediate levels with hollow, double-\textit{K}-vacancy configurations $2s2p$, $2s3p$, $2p3s$, $2p3d$, $2s4p$, $2p4s$ and $2p4d$  has been investigated. High-resolution photon-ion merged-beams measurements were carried out with the resolving power reaching up to 25,000  which is sufficient to separate the leading fine-structure components of the $2s2p~^3P$ term. Many-body perturbation theory was employed to determine level-to-level cross sections for  \textit{K}-shell excitation with subsequent autoionization. The resonance energies were calculated with inclusion of electron correlation and radiative contributions. Their uncertainties are estimated to be below $\pm 1$~meV. Detailed balance confirms the present photoionization cross-section results by comparison with previous dielectronic-recombination measurements. The high accuracy of the theoretical transition energies together with the present experimental results qualify photoabsorption resonances in heliumlike ions as new, greatly improved energy-reference standards at synchrotron radiation facilities.

\end{abstract}




\maketitle


\section{Introduction}
\label{Sec:introduction}
Structure and dynamics of the helium atom and heliumlike ions are among the most interesting subjects in fundamental atomic physics research. Two-electron atoms and ions are the simplest many-electron systems and thus, next to the hydrogen atom and the associated isoelectronic sequence, provide the best possible framework for theory to describe their physical properties with high accuracy. Yet, their three-particle nature and the competition between electron-electron and electron-nucleus interactions make them sufficiently complex to challenge detailed theoretical treatments~\cite{Tanner2000a}.

From an application-oriented point of view, two-electron ions provide an important tool for plasma diagnostics. Plasma parameters such as electron density and temperature can be derived from line-intensity ratios in the radiation emitted when heliumlike ions are present (\cite{Gabriel1969,Porquet2010,Kunze2009,Tallents2018} and references therein). This is especially important for astrophysics and fusion research.

Doubly excited states of helium, i.e., atoms with an empty $\textit{K}$ shell, are particularly intriguing. Such states were discovered by Compton and Boyce~\cite{Compton1928} by spectral analysis of the radiation emitted from a low-pressure helium discharge tube and then observed again by Kr\"{u}ger~\cite{Kruger1930} using a hollow-cathode discharge lamp. Both experiments found lines at wavelengths smaller than the minimum required for ionizing a helium atom and the authors discussed their assignment to $2\ell 2\ell' \to 1s2\ell$ transitions with $\ell, \ell' = s, p$. About thirty years later Madden and Codling~\cite{Madden1963} measured the absorption spectrum of helium by using synchrotron radiation and found unambiguous evidence for the population of doubly excited autoionizing levels. Their experiment initiated wide-spread and long-lasting interest in multiply excited states of atoms. Obviously, the doubly excited levels in the helium atom had been populated by absorption of a single photon,
\begin{equation}
\gamma + He(1s^2) \to He(n\ell n'\ell'),
\end{equation}
i.e., both $\textit{K}$-shell electrons had been excited simultaneously forming $n, n' \geq 2$ $^1P_1$ levels with energies sufficient to subsequently eject an electron via Auger decay. Subsequently, even the production of doubly excited $^3P$ levels by spin-forbidden absorption of a single photon by He$(1s^2~^1S)$ was observed (\cite{Penent2001,Argenti2008a} and references therein).

Photoexcitation producing doubly excited levels in helium atoms has been investigated in great detail~\cite{Domke1991a,Domke1992a,Domke1996,Puettner2001} by employing the increased photon flux available at second- and third-generation synchrotron light sources. Obtaining similar results for heliumlike atomic ions is much more demanding. The reason for this is the decrease of the pertinent cross sections with increasing atomic number and the difficulty to provide sufficiently high ion-target densities~\cite{Mueller1989}. In the experiments by Domke \textit{et al.}~\cite{Domke1996} He gas was used at a pressure in the range of 0.01-3~mbar which corresponds to a particle density greater than $2\times10^{14}$~cm$^{-3}$.

Photo-double excitation of heliumlike ions was observed in short-lived laser-produced plasmas for Li$^+$~\cite{Carroll1977} and Be$^{2+}(1s^2~^1S)$ ions~\cite{Jannitti1984}. The transient particle densities in these experiments reached  $10^{18}$~cm$^{-3}$ and higher. However, the photon energies that can be obtained with sufficient photon flux from a laser-produced plasma are limited. In experiments with quasi-static targets of atomic ions, the achievable densities are very much smaller but, depending on the light source employed, the photon energy range can go much beyond that of the plasma experiments. High-precision photo-single-excitation studies with heliumlike Fe$^{24+}$~\cite{Rudolph2013} and Kr$^{34+}$~\cite{Epp2015} employing an electron-beam ion trap (EBIT) were carried out with ion densities of a few $10^{10}$~cm$^{-3}$. Measurements employing targets of well-defined ions of given mass and charge are only possible when accelerated ion beams are employed in interacting-beams experiments~\cite{Mueller2008a}. In such beams  a typical upper limit of ion density is $10^6$~cm$^{-3}$. Therefore, photo-double excitation of  heliumlike ions in interacting-beams experiments has been restricted so far to Li$^+(1s^2~^1S)$~\cite{Scully2006a}. However, in contrast to the detailed experiments with neutral He and plasma-generated ions, the Li$^+$ photon-ion merged-beams experiment yielded independently-absolute cross sections for photoionization via doubly excited resonances.

There are numerous other ways to produce and to investigate doubly excited levels of heliumlike systems. These include double excitation of members of the helium isoelectronic sequence in collisions with atomic particles including electrons and solid foil targets (see, e.g., \cite{Mueller1989a,Kasthurirangan2013,Lyashchenko2017}), two-electron transfer in collisions of completely stripped ions with neutral particles (see, e.g., \cite{Mack1989}), photoionization of a lithiumlike atom accompanied by shake up, Auger decay of photon-induced triply excited lithiumlike states (see, e.g., \cite{Diehl1999a}), and dielectronic recombination of hydrogenlike ions (see, e.g., \cite{Kilgus1990,DeWitt1995,Bernhardt2011}). A discussion of all these experimental approaches is beyond the scope of this paper.

Theoretical approaches to the description of heliumlike systems and their physical properties have been reviewed by Tanner et al.~\cite{Tanner2000a}. Recent publications by Si \textit{et al.}~\cite{Si2016} and Goryaev \textit{et al.}~\cite{Goryaev2017} provide relatively accurate calculations of energy levels and transition rates of doubly excited states in heliumlike ions with atomic numbers $Z$ between 6 and 36. Si \textit{et al.}~\cite{Si2016} also list previous theoretical work on the subject.

In recent electron-ion collision experiments with heliumlike ions, doubly excited states were produced starting from the $1s2s~^3S_1$ levels of Li$^+$~\cite{Mueller2018b} and N$^{5+}$ ions~\cite{Mueller2014g}. With the parent ion already containing a $\textit{K}$-shell vacancy, excitation of a single electron can produce a doubly excited state with a relatively large cross section. A similar scheme is pursued in the present experiment. Here, photoionization of heliumlike metastable C$^{4+}$ ions,
 \begin{eqnarray}
 \label{Eq:CVPI}
\gamma + C^{4+}(1s2s~^3S_1)  \to ~ &   C^{4+}(n\ell n'\ell'~^3P_{0,1,2}) \nonumber \\
  & \downarrow\\
& C^{5+}(1s~^2S_{1/2}) + e, \nonumber
\end{eqnarray}
via intermediate doubly excited levels is studied both experimentally and theoretically  in the energy range of intermediate configurations $n\ell n'\ell' = 2s2p, 2s3p, 2p3s, 2p3d, 2s4p, 2p4s$, and $2p4d$. Similar experiments have only been reported for the neutral helium atom~\cite{Alagia2009}, but at a much lower resolving power compared to the present experiment and without providing absolute cross sections.

While the present work is part of the research effort dealing with the physics of heliumlike atoms and ions it is also part of an experimental endeavor to study $K$-shell photoexcitation and photoionization of carbon ions in different charge states. Investigations have been carried out previously on C$^{+}$~\cite{Schlachter2004a,Mueller2015a,Mueller2018a}, C$^{2+}$~\cite{Scully2005b}, and C$^{3+}$~\cite{Mueller2009a}. Photoionization of atomic ions via the production of a $K$-shell vacancy and the related aspects of precision measurements of the atomic structure of autoionizing states have been discussed in a recent review~\cite{Mueller2015c}. Since that review, new experimental investigations on photoionization of atomic ions involving $K$-shell electrons have been published for O$^{+}$, O$^{2+}$~\cite{Bizau2015},  O$^{4+}$,  O$^{5+}$~\cite{McLaughlin2017a}, Ne$^{+}$~\cite{Mueller2017}, O$^{-}$~\cite{Schippers2016a}, F$^{-}$~\cite{Mueller2018c}, Fe$^{20+}$, Fe$^{21+}$, Fe$^{22+}$, and Fe$^{23+}$\cite{Steinbruegge2015}.

This paper is organized as follows. After this introduction, Section~\ref{Sec:experiment} provides   a brief description of the experimental technique and the measurement of absolute photoionization cross sections. The energy calibration and related uncertainties are discussed in detail.  The calculations performed in the present context by employing many-body-perturbation theory with complex rotation with inclusion of contributions to the level energies arising from the quantization of the electromagnetic field are described in Section~\ref{Sec:theory}. In the main  section, \ref{Sec:results}, the experimental and theoretical results are presented. They are compared with one another and with information available in the literature. In particular, the principle of detailed balance is exploited to demonstrate the consistency of the present results on photoionization of C$^{4+}(1s2s~^3S)$ with previous measurements of absolute cross sections for dielectronic recombination of C$^{5+}(1s~^2S)$ ions. The paper ends with a summary and an outlook.

\section{Experiment}
\label{Sec:experiment}
The measurements were carried out using the PIPE (Photon-Ion Spectrometer at PETRA III) endstation of beamline P04~\cite{Viefhaus2013}  at one of the world's brightest synchrotron radiation sources, PETRA III, at DESY in Hamburg. The experimental setup and procedures have been described previously~\cite{Schippers2014,Mueller2017}. Photoionization of singly charged carbon ions employing PIPE has been reported~\cite{Mueller2015a,Mueller2018a}. Therefore, the description of  experimental details is restricted to issues directly related to the present C$^{4+}$ photoionization measurements.

\subsection{Experimental arrangement and procedures}
Heliumlike C$^{4+}$ ions were produced from methane gas in the hot plasma of an electron-cyclotron-resonance (ECR) ion source. The plasma chamber was set to a potential of +6~kV and the ions were extracted towards ground potential. With electrostatic lenses and steerers the ion beam was transported to the entrance aperture of a double-focusing 90-degree dipole bending magnet which separated the beam components according to their mass-over-charge ($m/q$) ratio. Behind the exit aperture of the magnet, the selected $^{12}$C$^{4+}$ ion beam component was transported further, again with electrostatic elements including a quadrupole triplet, parallel steering plates and a hemispherical 50-degree deflector, the merger,  which steered the ion beam onto the photon-beam axis and focused it to the center of a 50~cm long drift tube, the photon-ion merged-beam interaction region. With voltages of the order of 500~V on the drift tube, ions produced inside are energy-labeled~\cite{Schippers2014,Mueller2017} and thus, the length of the interaction region is defined. This is necessary for the measurement of absolute cross sections for photon-ion interactions.

The C$^{5+}$ photoionization products were separated from the C$^{4+}$ parent ion beam with a second double-focusing 90-degree  dipole bending magnet, the demerger. The parent ion beam was collected in a large Faraday cup located inside the magnet chamber while the product ions passed through apertures and were directed to a 180-degree out-of-plane hemispherical deflector which focused the product ion beam onto a single-particle detector. The out-of-plane deflection serves to suppress detector background arising from stray ions, electrons or photons. It cannot discriminate, though, against C$^{5+}$  ions produced in collisions with residual gas along their path between the merger and the demerger. Suppression of such background is only possible by keeping the vacuum pressure in that section as low as possible. In the interaction region a pressure of $3 \times 10^{-10}$~mbar could be maintained.

A further source of background is the occurrence of dark counts. The detector employed in the PIPE experiments is based on a well-tested geometry~\cite{Fricke1980a,Rinn1982} with a metallic ion-electron converter plate and a channel electron multiplier (CEM) amplifying the few-electron pulses released from the converter plate by impacting ions. 
All detector components are shielded inside a grounded metal box. The detection efficiency for atomic ions is close to 100\% ($0.97 \pm 0.03$).  By avoiding sharp edges of electrodes inside the box and by keeping the applied voltages  as low as possible (at most $\pm 1200$~V) the rate of dark counts has been reduced to the level of 0.02~s$^{-1}$.

The $^{12}$C$^{4+}$ ions have a $m/q$ ratio of approximately 3. Beside the desired carbon ions, the ion source also produced  H$_3^+$ with $m/q$ also close to 3. Closer inspection yields $m/q \approx 3.023~u/e$  for H$_3^+$ where $u$ is the atomic mass unit and $e$ the elementary charge. For $^{12}$C$^{4+}$ $m/q \approx 2.999~u/e$. The 8 per mille difference is by far enough to separate the different ion species by the first dipole magnet.

Magnetic separation of  $^{12}$C$^{4+}(1s^2~^1S_0)$ ground-state ions from metastable $^{12}$C$^{4+}(1s2s~^3S_1)$ is not possible. From previous experiments on electron-impact ionization of heliumlike ions it is known that the ECR ion source produces $^3S$ metastable ions while the population of $^1S$ ions is negligibly small~\cite{Mueller2014g,Mueller2018b}. This can be rationalized by the different lifetimes $\tau$ (20.59~ms~\cite{Schmidt1994} for C$^{4+}(1s2s~^3S_1)$ and only 3.03~$\mu$s~\cite{Derevianko1997} for C$^{4+}(1s2s~^1S_0)$) and the different statistical weights of the $^3S$ and $^1S$ levels. Most important in this context is the approximate flight time of about 15~$\mu$s of $^{12}$C$^{4+}$ ions from the source to the interaction region leaving little room for the survival of C$^{4+}(1s2s~^1S_0)$) excited levels. Hence, the parent C$^{4+}$  beam consisted of ions in the $1s^2~^1S$ ground level with a fraction $1-f$ and the $1s2s~^3S$ metastable level with a fraction $f$. From previous experiments in which the same type of ion source was employed, it is known that $f$ was 6\% for a beam of N$^{5+}$ ions~\cite{Mueller2014g} and 13.6\% for a beam of Li$^+$ ions~\cite{Mueller2018b}. The fraction of C$^{4+}(1s2s~^3S_1)$ had to be expected to be somewhere between these two numbers. The exact value is to be determined by comparison with theory and other experiments (see below).

For the measurement of absolute cross sections the ion beam was strongly collimated by variable apertures in front of and behind the interaction region. With these apertures closed to about 1~mm $\times$ 1~mm and almost touching the photon beam, the ion beam was optimized for transmission and, by that, very good overlap with the photon beam was enforced. Then the aperture behind the interaction region was opened to a size 3~mm $\times$ 3~mm and the beam overlap factor was measured with six independent slit scanners yielding values of around 4400~cm$^{-1}$. The ion current in the absolute measurements was about 1~nA. At 359~eV the photon flux was $4.2 \times 10^{11}$~s$^{-1}$ at a bandwidth of 16.2~meV. Under these conditions the background count rate of C$^{5+}$ ions produced by electron-stripping collisions with residual gas components was about 2~s$^{-1}$. The maximum signal count rate on the $2s2p~^3P_2$ resonance was slightly below 4~s$^{-1}$. Unfortunately, the dominant ($\approx$90\% ground-level) fraction of the parent beam contributed to the electron-stripping background but not to the $^3P$ resonance signals which can exclusively be reached from the $^3S$ metastable component of the parent beam. As it turned out, only about 10\% of the total C$^{4+}$ beam contributed to the measured signal.

Without considering the uncertainty arising from the determination of the metastable-ion fraction $f$, the apparent cross sections measured with a mixed beam of ground-state and metastable C$^{4+}$ ions have a systematic uncertainty of 15\%~\cite{Schippers2014}. The apparent cross sections were normalized to 100\% parent metastable ions by multiplication with $f^{-1}$. The fraction $f$ was determined by comparison with the results of the present theory. By employing the principle of detailed balance photoionization cross sections can be converted to photorecombination cross sections. Comparison with absolute experimental data for dielectronic recombination of C$^{5+}(1s)$ ions, obtained at a heavy-ion storage ring, shows excellent agreement (within 3\%, see below) thus confirming the present determination of $f$.

\subsection{Energy calibration and related uncertainties}
The photon energy was calibrated against the position of the $2p \to 4s$ excitation resonance in neutral Ar at 244.39~eV and the position of the lowest vibrational level ($\nu = 0$) reached in the N$1s \to \pi*$ excitation of the  neutral N$_2$ molecule at 400.88~eV. The reference energies originate from electron-energy-loss spectroscopy (EELS) experiments~\cite{King1977a,King1977b,Hitchcock1980a,Sodhi1984a,Kato2007,Ren2011} for which low uncertainties in the energy determination are quoted. However, one has to keep in mind that quoted uncertainties tend to be optimistic. In a previous detailed study of characteristic energies in neutral Ne and Ne$^+$~\cite{Mueller2017} numerous experimental results were compared with one another. Discrepancies between different experiments much larger than the quoted uncertainties were found which casts doubt on the low error bars quoted in the literature. In the case of the Ne $1s \to 3p$ transition energy at 867.29~eV a realistic uncertainty is presently 0.2~eV in contrast to quoted error bars as low as 20~meV.

Reference energies relevant to the present calibration are
\begin{itemize}
   \item  for the Ar ($2p \to 4s$) transition
          \begin{itemize}
           \item
           $244.39 \pm 0.01$~eV~\cite{King1977a}
           \item
           $244.37 \pm 0.02$~eV~\cite{Sodhi1984a}
           \item
           $244.390 \pm 0.004$~eV~\cite{Ren2011}
           \end{itemize}
   \item  for the N$_2$ (N$1s \to \pi*,\nu = 0$) transition
          \begin{itemize}
           \item
           $400.86 \pm 0.03$~eV~\cite{King1977b}
           \item
           $400.70 \pm 0.05$~eV~\cite{Hitchcock1980a}
           \item
           $400.88 \pm 0.02$~eV~\cite{Sodhi1984a}
           \item
           $400.865 \pm 0.02$~eV~\cite{Kato2007} (relative to the Ar ($2p \to 4s$) transition at 244.39~eV)
           \end{itemize}
\end{itemize}
 The energy determination in EELS experiments is directly associated with the measurement of the voltage on the electron spectrometer. Thus the accuracy of transition energies from EELS experiments critically depends on the sensitivity and accuracy of the voltmeter used. The EELS experiments on the  Ar calibration line agree with one another within their quoted uncertainties. This is not the case for the N$_2$ reference.  Discrepancies have been explained by the different accuracies of digital voltmeters used in the different experiments. However, deficiencies in the performance of the voltmeter should be reflected in the quoted uncertainty of a measured transition energy. Obviously, this was not the case in the experiments by Hitchcock and Brion~\cite{Hitchcock1980a} and Sodhi and Brion~\cite{Sodhi1984a} whose numbers differ by 180~meV while uncertainties of only 50~meV and 20~meV have been quoted. Thus, it cannot be excluded that the error bar on the 400.88~eV N$_2$ (N$1s \to \pi*,\nu = 0$) transition energy quoted by Sodhi and Brion~\cite{Sodhi1984a} is greater than 20~meV.

The situation calls for improved calibration standards for synchrotron-radiation experiments in the photon-energy range from approximately 300~eV to 1~keV. The uncertainties of the existing standards are so large, that state-of-the-art atomic structure theory can not at all be tested for light few-electron atomic systems by using synchrotron radiation in that energy range. 
 A possible way to deal with this problem  is to calibrate experiments by comparing measured resonance positions with accurate theoretical energies. In the energy range of interest, moderately highly charged ions of light elements have to be considered. These are not typically available at synchrotron radiation sources. However, by using the PIPE setup it would be possible to transfer high-quality calibration standards to the gas-phase standards that are presently in use at synchrotrons world wide.

In a recent publication, level energies for $1s^2 2\ell$ and $1s 2\ell 2\ell'$  states in lithiumlike ions with atomic numbers $Z$ between 6 and 17 have been presented~\cite{Yerokhin2017a,Yerokhin2017b} with uncertainties of a few meV, much smaller than the typical calibration uncertainties of common neutral-gas standards. Transitions in lithiumlike C$^{3+}$ through Ne$^{7+}$ would adequately cover the energy range of interest with resonances at about 300~eV to more than 900~eV. These ions can be produced with the ECR source presently installed at PIPE so that a recalibration of gas standards is now possible. Also the present investigation with heliumlike C$^{4+}$ ions has the potential for providing new calibration standards.

For comparison of the present experimental data with the present theoretical calculations it is desirable to have a theory-independent calibration. Therefore, the ``conventional'' calibration procedure is discussed further. Based on past experience it was assumed that the deviation between nominal energies provided by the beamline and the true photon energy is a linear function of the nominal photon energy. In order to determine this linear function the Ar and N$_2$ resonance energies mentioned above were employed. With the uncertainties of the reference energies, the linear function has uncertainties which increase when it is extrapolated to energies beyond 400~eV. By assuming an increased uncertainty of 30 instead of 20~meV for the N$_2$ (N$1s \to \pi*,\nu = 0$) transition found at the uncertainty-weighted average energy of 400.86~eV that results from the existing measurements~\cite{King1977b,Hitchcock1980a,Sodhi1984a,Kato2007}, the possible error of the energy axis set by the calibration standards is 40~meV at 440~eV.

There are additional sources of uncertainty in the present calibration for the measurements with C$^{4+}$ ions. Since the ions move against the direction of the photon beam the resulting Doppler shift has to be corrected for. The Doppler-corrected energy $E_\mathrm{D}$ for interacting ion and photon beams determined from the photon energy $E_\mathrm{Lab}$ in the laboratory frame is:
\begin{equation}
E_\mathrm{D} =  \frac{E_\mathrm{Lab}}{\gamma (1+\beta \mathrm{cos}\theta)},
\end{equation}
with the Lorentz factor
\begin{equation}
\gamma = \frac{1}{\sqrt{1-\beta^2}},
\end{equation}
and the angle $\theta$ between the two beam directions. The ion velocity $v = \beta c$ where $c$ is the vacuum speed of light can be inferred from $\gamma$ which is related to the kinetic energy of the ions,
\begin{equation}
E_\mathrm{kin} = q e U_\mathrm{acc} = (\gamma-1)m_{\mathrm{i}0}c^2.
\end{equation}
Here, $q$ is the charge state of the parent ions, $e$ the elementary charge, $U_\mathrm{acc}$ the ion acceleration voltage, and $m_{\mathrm{i}0}$ the rest mass of the ion. Thus,
$\beta$ can be calculated as
\begin{equation}
\beta = \sqrt{1-\frac{1}{(1+x)^2}},
\end{equation}
with $x$ defined as $q e U_\mathrm{acc}/(m_{\mathrm{i}0}c^2)$. The Doppler-corrected photon energy is then given by
\begin{equation}
E_\mathrm{D} = \frac{E_\mathrm{Lab}}{(1+x) + \sqrt{2x+x^2} \, \mathrm{cos}\theta}.
\end{equation}
For counter-propagating beams the angle is $\theta = 180^\circ$ and $\mathrm{cos} \theta = -1$. This results in
\begin{equation}
E_\mathrm{D} = \left[ (1+x) + \sqrt{x} \sqrt{2+x} \, \right] E_\mathrm{Lab}.
\end{equation}

Uncertainty arises from the unknown plasma potential in the ion source which changes the effective acceleration voltage. The plasma potential in an ECR ion source is expected to be no more than 50~V~\cite{Tarvainen2004}. Thus, the uncertainty of the Doppler correction is at most about 3~meV at $E_\mathrm{Lab} = 360$~eV in the present case. Assuming that the potential of the ion source plasma chamber was measured with an uncertainty smaller than 0.5\% an additional possible error of 2~meV results. If the angle $\theta$ between the two beams deviates from 180 degrees, the Doppler correction changes accordingly. Given the tight collimation of the ion beam with $\pm 0.6$~mm at the entrance to the interaction region and $\pm 1.1$~mm at the exit, the maximum deviation in the angle $\theta$ is 3.4~mrad corresponding to a shift of about  2~meV  at 360~eV.

Another source of uncertainty is the stability of the photon source geometry which is taken into account in the calculation and the control of the actual photon energy in real time. Depending on temperature and operation mode of the synchrotron ring the electron beam may change its position in the ring. Changes in the position of the stored electron beam have an immediate impact on the photon energy transported to the experiment. Despite the tremendous achievements in beam position stability, drifts of several meV within a few hours of seemingly stable ring and monochromator operation have been observed. Moreover, the carefully calibrated in-vacuum angular encoders of both pre-mirror and grating  introduce an uncertainty which has been reduced to a level of about 10~ppm which corresponds to 4 meV at 400~eV. During a period of one top-up ring-filling cycle the electron beam can move adding another 1~meV periodic energy shift. In summary,  calibration uncertainties (one standard deviation) of the present measurements are estimated to be 40~meV at 360~eV and 50~meV  at 440~eV.

\begin{table*}
\caption{\label{tab:resonancedetails} Contributions to the calculated total energies of the fine-structure components of C$^{4+}(2s2p~^3P)$ and the lowest-energy triplet state C$^{4+}(1s2s~^3S_1)$. The results are given in atomic units  for $^{12}C$. To convert to eV multiply with $E_\mathrm{h} \times M/(m_\mathrm{e} + M)$ where $E_\mathrm{h}$ is the Hartree energy in eV,  $m_\mathrm{e}$ is the electron rest mass and $M$ the rest mass of the $^{12}$C nucleus. The factor $M/(m_\mathrm{e} + M)$ accounts for the normal mass shift.
 With $E_\mathrm{h} = 27.211~386~02(17)$~eV and $m_\mathrm{e} = 5.485~799~090~70(16) \times 10^{-4}$~u according to the 2014 CODATA (Committee on Data for Science and Technology) recommended values~\cite{CODATA2014}, the conversion factor is  27.210~141~77~eV.
}
\begin{ruledtabular}
\begin{tabular}{llllr}
& $2s2p~^3P_0$ & $2s2p~^3P_1$ & $2s2p~^3P_2$ & $1s2s~^3S_1$\\
\hline
Coulomb interaction  $1^{st}$ order & -8.207838 & -8.207217 & -8.205970 & -21.118982\\
Breit interaction $1^{st}$ order  & ~0.000075 & ~0.000011 & ~0.000019 & ~0.000569\\
$\Delta$ Coulomb and Breit  all orders
& -0.027206 & -0.027193 & -0.027188 & -0.312848\\
$\Delta$ One-electron radiative corrections
& ~0.000114 & ~0.000116 & ~0.000119 & ~0.001009\\
$\Delta$ Two-electron radiative corrections
& -0.000006 & -0.000007 & -0.000009 & -0.000032\\
\hline
Total & -8.234861 & -8.234290 & -8.233030  & -21.430284\\
Drake~\cite{Drake1988,Johnson1985a} & & & & -21.430301\\
\end{tabular}
\end{ruledtabular}
\end{table*}

\section{Theory}
\label{Sec:theory}
In the present theoretical treatment the energies of the doubly and singly excited states are calculated with relativistic many-body perturbation theory in an all-order formulation including  single and double excitations,  as described by Salomonson and \"{O}ster~\cite{Salomonson1989a}. This means that  all types of excitations that can be formed in a pure two-electron system are accounted for. The  C$^{4+}$ ion is placed in a spherical
box within which a discrete radial grid is used. Diagonalization of the  discretized hydrogenlike Dirac Hamiltonian gives a discrete basis set, complete on the grid chosen. The basis set is then used to construct correlated wave functions to all orders in the perturbation expansion of the  electron-electron interaction.  Here, both  the Coulomb  and the Breit interaction are accounted for. The perturbation expansion is constructed from an {\em extended model space}~\cite{Lindgren1974} whenever a state is dominated by two or more nearly degenerate configurations. An example is  the $2s2p~^3P_1$ state which in $jj$-coupling has major contributions both from the  $2s2p_{1/2}$ and  the $2s2p_{3/2}$ configuration. This is a common scenario in $jj$-coupling, but also for doubly excited states in general. A multipole expansion of the electron-electron interaction is used, making the method applicable to many-electron atoms in general. The present calculations include all contributing partial waves up to $\ell_{max} =10$.

When perturbation theory is applied to autoionizing states it is obvious that the use of a discrete basis set will cause problems close to the poles in the energy denominator. A complex scaling of the radial coordinates
can, however, solve this problem.  The present treatment follows the method employed by Lindroth~\cite{Lindroth1994a} for the calculation of doubly excited levels in the helium atom, and later for a number of Be-like ions (see e.g.~\cite{Lestinsky2008a}). The method yields complex energies for  the autoionizing states, where the imaginary part corresponds to  the half-life time (due to Coulombic decay) of the state. The  decay rates due to photon emission are calculated from the dipole matrix elements between the  doubly excited states and the $\left(1sn\ell\right)~^3L_J$ states with $n \leq 4$. The radiative decay rates of all the considered doubly excited states are completely dominated (with contributions of more than 99\%)  by photo-emission events that require one-electron transitions only.

For the  underlying theory of complex rotation (CR) and many-body-perturbation theory (MBPT) the reader is referred to a review by Lindroth and Argenti~(\cite{Lindroth2012a} and references therein).

The additional contributions to the energies originating from the quantization of the electromagnetic field  are treated with the procedure implemented in the MCDFGME (Multi Configuration Dirac Fock and General Matrix Element) code, developed by Desclaux and Indelicato~\cite{Desclaux1975,Indelicato1990,Indelicato1987,Indelicato2005}. It includes one-electron one-loop corrections (self energy ~\cite{Mohr1974,Mohr1992a,Mohr1992b,Indelicato1992b,LeBigot2001} and vacuum polarization), two-loop one-electron corrections (two loop self-energy, mixed self-energy and vacuum polarization diagrams, K\"{a}ll\'{e}n and Sabry potential contributions, see Ref. \cite{Yerokhin2008} and references therein) in the Coulomb field of the nucleus, although the result is almost completely (99\%) determined  by the one-photon self energy and the one-loop vacuum polarization. It also includes vacuum polarization due to the electronic potential, retardation beyond the Breit interaction and the effect from the electron-electron interaction on the self-energy evaluated  with the so called Welton method~\cite{Welton1948,Indelicato1987,Indelicato1990}. The latter method has recently been tested against the model operator approach developed by Shabaev and collaborators \cite{Shabaev2013,Shabaev2015} for $Z=18$~\cite{Machado2018a} and relative differences of less than $0.01$\% were found.

The normal mass shift is taken care of by the correction factor $M/(m_\mathrm{e} + M)$ where $m_\mathrm{e}$ is the electron rest mass and $M$ the rest mass of the $^{12}$C nucleus.  The specific mass shift has also been considered. Calculations of the mass-polarization effect with the non-relativistic formula resulted in shifts of at most 0.1~meV. It is safe to say that mass-polarization shifts of the investigated levels of the heliumlike C$^{4+}$ ion are well below the 1~meV level.

Table~\ref{tab:resonancedetails} illustrates the importance of different contributions to the level energies of the $2s2p~^3P_J$ fine structure components. The  C$^{4+}$ ion has a sufficiently low atomic number, $Z=6$, so that Coulomb correlation dominates by many orders of magnitude over the Breit interaction and the radiative contributions. The one- and two-particle radiative corrections are given separately showing that the latter, which are less well known, contribute with only a few tenths of a  meV to the transition energies and thus do not affect the comparison between theory and experiment on a measurable level. The contributions to the
$1s2s~^3S_1$ state, the lowest of the triplets, are also given in Table~\ref{tab:resonancedetails}.

The present calculation can be compared to the results obtained previously by  Drake~\cite{Drake1988}. He used highly correlated non-relativistic wave functions of Hylleraas type to calculate the ionization energy. Relativistic and radiative corrections are subsequently treated as perturbations. The calculation is thus very different from the one presented here. Yet,  the two calculations agree to within $0.5$~meV.  The number  at the bottom of
Table~\ref{tab:resonancedetails} is obtained by adding Drake's result for the ionization energy  to the well-established
values for hydrogenlike systems by Johnson and Soff~\cite{Johnson1985a}.

\section{Results}
\label{Sec:results}

\begin{table*}
\caption{\label{tab:resonanceparameters} Calculated parameters of the  21 lowest resonance contributions to single photoionization of metastable C$^{4+}(1s2s~^3S)$. All excited levels are associated with $^3P$ terms due to the selection rules for electric dipole transitions. The columns provide the configurations, the total angular momenta $J$, the resonance energies $E_\mathrm{res}$ relative to the parent ion, the natural (life-time) widths $\Gamma$, the Fano $q$ parameters~\cite{Fano1961}, the absorption resonance strengths $S_\mathrm{abs}$, the Auger decay rate $A_\mathrm{a}$, the total radiative rate $A_\mathrm{r}$, the branching ratio for Auger decay $B_\mathrm{a}$, and the resulting ionization resonance strengths $S_\mathrm{ion}$. Numbers in square brackets are powers of 10. For further details see text.}
\begin{ruledtabular}
\begin{tabular}{cccccccccc}
config. &  $J$  & $E_{\mathrm res}$ &  $\Gamma$     & $q$         &$S_\mathrm{abs}$ &$A_\mathrm{a}$ & $A_\mathrm{r}$ &  $B_\mathrm{a}$ &  $S_\mathrm{ion}$  \\
        &       &      eV       &       eV      &                 &   Mb\,eV    & $10^{13}$~s$^{-1}$ & $10^{11}$~s$^{-1}$ & &     Mb\,eV  \\
\hline
2s2p	&	0	&	359.0493	&	9.229[-03]	&	-319.9	      &	4.606[+00]	&	1.402[+00]	&	7.058	&	9.521[-01]	&	4.385[+00]\\
     	&	1	&	359.0649	&	9.053[-03]	&	-320.3	      &	1.382[+01]	&	1.375[+00]	&	7.058	&	9.512[-01]	&	1.314[+01]\\
     	&	2	&	359.0992	&	8.897[-03]	&	-321.2	      &	2.303[+01]	&	1.352[+00]	&	7.059	&	9.504[-01]	&	2.188[+01]\\
2s3p	&	0	&	418.7346	&	3.898[-03]	&	-132.6	      &	3.097[-01]	&	5.923[-01]	&	4.744	&	9.258[-01]	&	2.867[-01]\\
     	&	1	&	418.7455	&	3.842[-03]	&	-137.7	      &	9.734[-01]	&	5.837[-01]	&	4.650	&	9.262[-01]	&	9.016[-01]\\
     	&	2	&	418.7658	&	3.815[-03]	&	-148.3	      &	1.771[+00]	&	5.796[-01]	&	4.468	&	9.284[-01]	&	1.644[+00]\\
2p3s	&	0	&	419.1274	&	1.532[-04]	&	-3530	      &	4.371[-01]	&	2.328[-02]	&	6.194	&	2.732[-01]	&	1.194[-01]\\
     	&	1	&	419.1373	&	1.295[-04]	&	-4147	      &	1.265[+00]	&	1.967[-02]	&	6.293	&	2.381[-01]	&	3.013[-01]\\
     	&	2	&	419.1602	&	1.260[-04]	&	-6027	      &	1.955[+00]	&	1.914[-02]	&	6.488	&	2.278[-01]	&	4.454[-01]\\
2p3d	&	0	&	423.3094	&	2.493[-07]	&	1.691[+04]	  &	1.137[-01]	&	3.788[-05]	&	6.687	&	5.662[-04]	&	6.439[-05]\\
     	&	1	&	423.3029	&	2.208[-06]	&	1.725[+04]	  &	3.434[-01]	&	3.354[-04]	&	6.682	&	4.995[-03]	&	1.715[-03]\\
     	&	2	&	423.2922	&	3.608[-07]	&	1.734[+04]	  &	5.802[-01]	&	5.481[-05]	&	6.672	&	8.209[-04]	&	4.763[-04]\\
2s4p	&	0	&	437.5709	&	1.556[-03]	&	-156.1	      &	1.269[-01]	&	2.364[-01]	&	4.339	&	8.450[-01]	&	1.072[-01]\\
     	&	1	&	437.5797	&	1.538[-03]	&	-160.3	      &	3.980[-01]	&	2.337[-01]	&	4.220	&	8.470[-01]	&	3.371[-01]\\
     	&	2	&	437.5966	&	1.532[-03]	&	-168.3	      &	7.202[-01]	&	2.328[-01]	&	4.000	&	8.533[-01]	&	6.145[-01]\\
2p4s	&	0	&	438.0040	&	9.150[-05]	&	-971.0	      &	1.343[-01]	&	1.390[-02]	&	6.092	&	1.858[-01]	&	2.495[-02]\\
     	&	1	&	438.0125	&	7.921[-05]	&	-1045	      &	3.837[-01]	&	1.203[-02]	&	6.223	&	1.621[-01]	&	6.217[-02]\\
     	&	2	&	438.0301	&	5.057[-05]	&	-1254	      &	5.763[-01]	&	7.683[-03]	&	6.482	&	1.060[-01]	&	6.107[-02]\\
2p4d	&	0	&	439.5723	&	1.573[-07]	&	-3.646[+04]	  &	4.526[-02]	&	2.390[-05]	&	6.329	&	3.775[-04]	&	1.708[-05]\\
     	&	1	&	439.5664	&	5.681[-06]	&	-2.850[+04]   &	1.377[-01]	&	8.631[-04]	&	6.314	&	1.348[-02]	&	1.856[-03]\\
     	&	2	&	439.5559	&	1.711[-07]	&	-2.576[+04]	  &	2.363[-01]	&	2.599[-05]	&	6.287	&	4.132[-04]	&	9.763[-05]\\
\end{tabular}
\end{ruledtabular}

\end{table*}
The main results of the present theoretical approach are displayed in Table~\ref{tab:resonanceparameters}. Calculations were carried out for all levels within the configurations $2s2p, 2s3p, 2p3s, 2p3d, 2s4p, 2p4s$, and $2p4d$ that can be reached by electric-dipole excitation from the metastable C$^{4+}(1s2s~^3S_1)$ level. Only $^3P$ excited levels with total angular momenta $J = 0, 1, 2$ are dipole allowed. Hence, within the selected configurations 21 levels had to be considered for calculating the cross sections for single photoionization of C$^{4+}(1s2s~^3S_1)$. These resonances were expected to provide sizable contributions to the total ionization which also includes direct removal of the $2s$ electron from the metastable $1s2s~^3S$ level. The threshold for  direct $2s$ ionization is 93.131~eV~\cite{NIST2018} where the cross section $\sigma_{2s}$ is about 0.57~Mb. In the energy range of present interest, 350 to 450~eV, $\sigma_{2s}$ drops from 0.0306~Mb to 0.0159~Mb. Removal of the $1s$ electron from metastable C$^{4+}(1s2s~^3S)$ does not occur at energies lower than 460.608~eV~\cite{NIST2018}.

The entries in Table~\ref{tab:resonanceparameters} include the information defining the spectroscopic notation for each of the resonances. As an example, the first row in the table is for the $2s2p~^3P_0$ resonance. For the calculation of resonance cross sections the resonance parameters have to be known, i.e.,  the resonance energy $E_\mathrm{res}$, the natural width $\Gamma$, the Fano asymmetry parameter $q$~\cite{Fano1961}, and the resonance strength $S$. To determine the contribution of a given resonance to the photoabsorption cross section, knowledge about the associated strength $S_\mathrm{abs}$ is required. It is provided in the sixth column of Table~\ref{tab:resonanceparameters}. Resonant photoabsorption determines the total intermediate population of the resonant level. This level can decay by autoionization or by photoemission. In the present experiments, the observed final channel was that of net single ionization of C$^{4+}$, i.e., the C$^{5+}$ final products were registered. To obtain the ionization cross section, the relative probability for Auger decay has to be known. This branching ratio follows from
\begin{equation}
\label{Eq:branchingratio}
B_\mathrm{a} = \frac{A_\mathrm{a}}{A_\mathrm{a}+A_\mathrm{r}},
\end{equation}
where, in the present context, $A_\mathrm{a}$ is the Auger decay rate from the specified intermediate doubly excited level to C$^{5+}(1s) + e$, and  $A_\mathrm{r}$ is the radiative-decay rate of the intermediate doubly excited level to all levels of the type C$^{4+}(1sn\ell)$ with $n \leq 4$  included in the calculation. The radiative decay rates  have been calculated in first order perturbation theory. The possibility that neighbouring resonances can affect each other, as recently discussed in Refs.~\cite{Beyer2017a,Horbatsch2010} has not been considered. The parameters thus obtained are provided in Table~\ref{tab:resonanceparameters}. The resonance strength for ionization is given by
\begin{equation}
S_\mathrm{ion} = B_\mathrm{a} S_\mathrm{abs}.
\end{equation}
Thus, the cross section $\sigma_\mathrm{PI}$ for photoionization of C$^{4+}(1s2s~^3S_1)$ in the photon energy range 350 to 450~eV can be represented as~\cite{Schippers2002a}:
\begin{equation}
\sigma_\mathrm{PI} = \sigma_{2s} + \sum^{21}_{k=1}{\frac{2 S_{\mathrm{ion},k}}{\pi q^2_k \Gamma_k}\left[\frac{(q_k + \epsilon_k)^2}{(1+\epsilon_k^2)}-1\right]},
\end{equation}
that is, a sum over all intermediate autoionizing doubly excited levels characterized by their individual resonance parameters which are labeled with the resonance number $k = 1, 2, ...,21$. The reduced energy $\epsilon_k$ of the $k^\mathrm{th}$ resonance is given by
\begin{equation}
\epsilon_k(E) = \frac{2 (E-E_{{\mathrm res},k})}{\Gamma_k}.
\end{equation}

\begin{figure}
\includegraphics[width=\columnwidth]{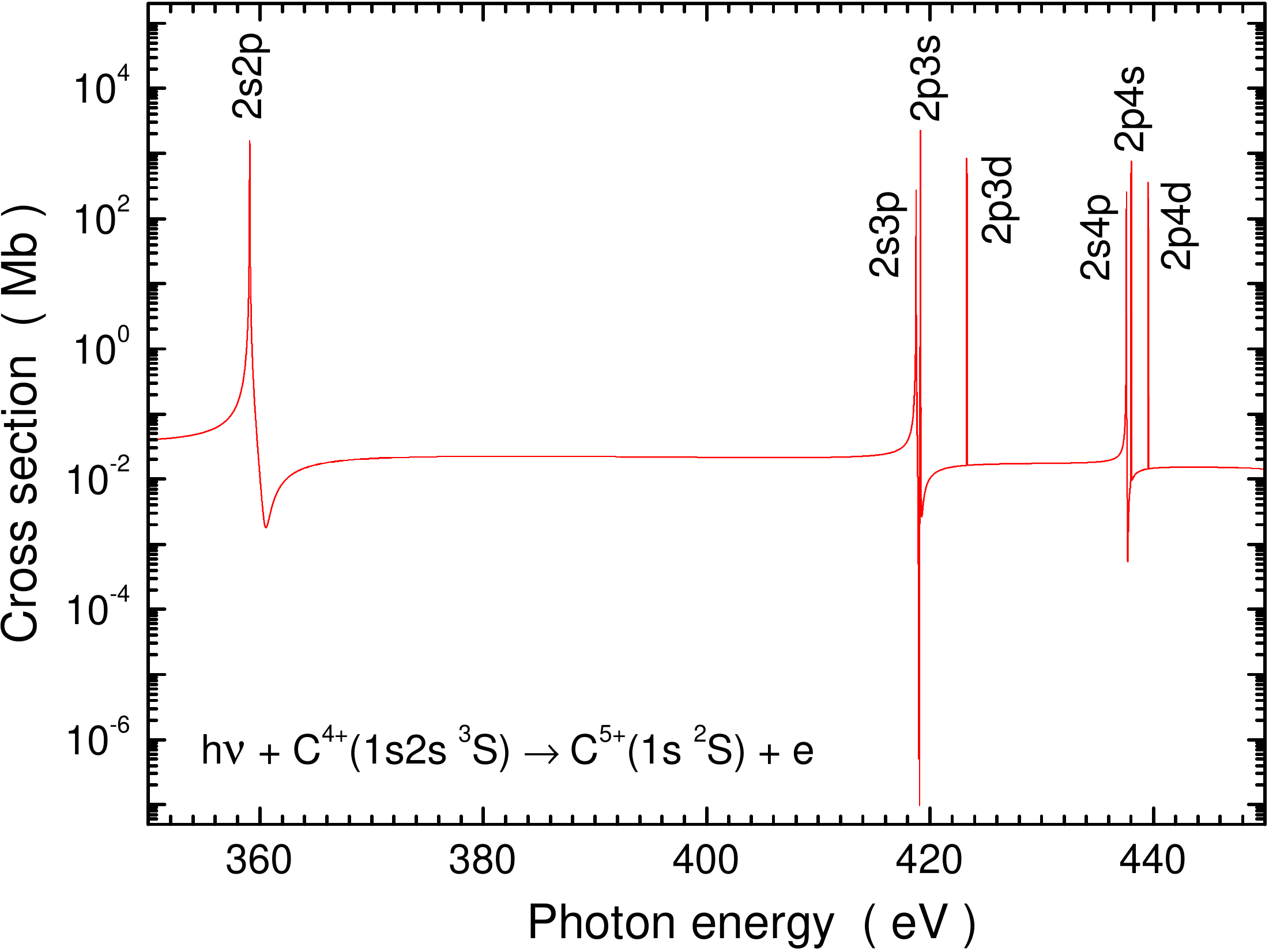}
\caption{\label{Fig:theory} (color online) Calculated cross sections for photoionization of C$^{4+}(1s2s~^3S)$ ions in the energy range of intermediate doubly excited $2s2p$, $2s3p$, $2p3s$, $2p3d$, $2s4p$, $2p4s$, and  $2p4d$ configurations. No convolution with an experimental response function was applied.
}
\end{figure}

\begin{figure*}
\includegraphics[width=17.9cm]{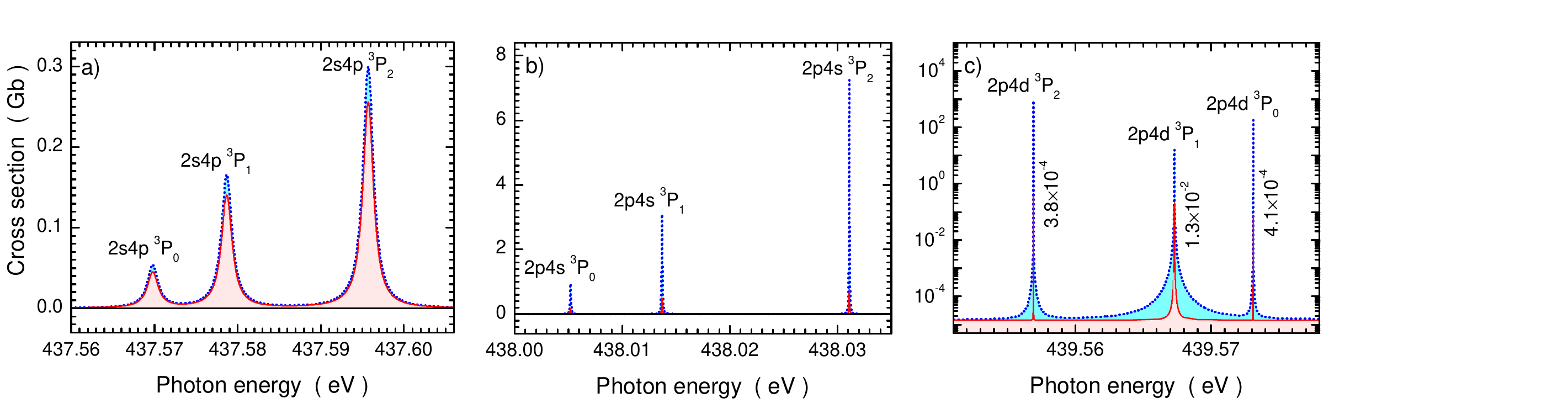}
\caption{\label{Fig:branching} (color online) Calculated natural (unconvoluted) cross sections for photoabsorption (dotted line with  light blue shading) and photoionization (solid line with light red shading) of C$^{4+}(1s2s~^3S)$ ions via intermediate a) $2s4p~^3P$, b) $2p4s~^3P$, and c) $2p4d~^3P$ levels which then decay by the emission of photons or electrons, respectively. Note the logarithmic scale in panel c) which was necessary to illustrate the small Auger branching ratios provided separately as numbers for each of the three $2p4d~^3P_{0,1,2}$ resonances.
}
\end{figure*}

Figure~\ref{Fig:theory} shows the theoretical cross section function in the energy range 350 to 450~eV. Seven groups of resonances are identified by their electronic configurations. The natural (unconvoluted) cross section covers a very large range from about $10^{-6}$ to more than 1500~Mb. This can only be made visible in a logarithmic plot which optically overemphasizes the small direct-ionization cross section $\sigma_{2s}$. Destructive interference of the resonant and direct ionization pathways results in deep but narrow dips in the cross-section function. The natural widths range from $1.57 \times 10^{-7}$ to $9.23 \times 10^{-3}$~eV spanning almost five orders of magnitude. Clearly, such narrow features cannot be resolved in an overview plot like the present one.

For the illustration of details within the groups of $2\ell n\ell'$ resonances, Fig.~\ref{Fig:branching} shows the natural (unconvoluted) cross sections of the $2s4p$ resonance levels (panel a), of the $2p4s$ resonance levels (panel b) and the $2p4d$ resonance levels (panel c). The cross sections for photoabsorption and for photoionization are shown by dotted and solid lines, respectively. According to Eq.~\ref{Eq:branchingratio} they differ by the factor $B_\mathrm{a}$, the branching ratio for autoionization. For doubly excited states of low-$Z$ ions, it is common knowledge that Auger decay probabilities typically exceed radiative decay rates by far. This is confirmed by Fig.~\ref{Fig:branching}a for the $2s4p$ resonance group, where the branching ratios $B_\mathrm{a}$ are approximately 0.85 (see Table~\ref{tab:resonanceparameters}, 9$^\mathrm{th}$ column, where the dominance of Auger over radiative decay becomes all the more obvious for the $2s2p$ and $2s3p$ resonances). However, the situation is very much different for the $2p4s$ (as well as $2p3s$) resonances where $B_\mathrm{a}$ is between 0.10 and 0.19 indicating that radiative stabilization now dominates the decay. For the $2p4d$ (and a little less pronounced also for $2p3d$) resonances this effect is dramatic. Branching ratios $B_\mathrm{a}$ between 0.014 and 0.00037 are predicted. Once populated, the $2p4d~^3P$ autoionizing levels decay almost exclusively by emission of photons and therefore, they contribute little to the ionization cross section. This is the reason why the first photospectroscopy experiments producing doubly excited levels of helium could actually see a signal. It also provides a rationale for the surprise result of Kasthurirangan \textit{et al.}~\cite{Kasthurirangan2013} who saw photon emission from doubly excited $2p3d~^1P$ heliumlike Si, S, and Cl ions.

It should be mentioned in the context of Fig.~\ref{Fig:branching} that experimentally resolving the individual fine-structure components that are clearly separated in the theory plots  would require a resolving power $E/\Delta E$ greater than 50,000 for the $2s4p$ resonances and  greater than 100,000 for the $2p4d$ resonances. This is presently not experimentally achievable. Moreover, with realistic bandwidths of 100~meV available at about 438~eV the narrow $2p4d$ resonances would produce apparent photoionization cross sections of the order of 0.1~Mb, far too small to be seen in the present experiments given the relatively high background and the small flux of metastable C$^{4+}$ parent ions available.

\begin{table*}
\caption{\label{tab:comptheories} Comparison of results of the present calculations with theoretical data obtained by Goryaev \textit{et al.}~\cite{Goryaev2017}. The first two columns provide the configurations and the total angular momenta $J$ of the $^3P$ terms which are accessible by photoexcitation of C$^{4+}(1s2s~^3S)$. The quantities that can be compared are the level energies $E$ relative to the C$^{4+}(1s2s~^3S)$ initial state, the Auger rates $A_\mathrm{a}$ and the total radiative rates $A_\mathrm{r}$. The present theory data are marked by the superscript ``this work'', those of Goryaev, Vainshtein and Urnov~\cite{Goryaev2017} by the superscript ``GVU''. Numbers in square brackets are powers of 10.
}
\begin{ruledtabular}
\begin{tabular}{cccccccc}
config. & J  & $E^\mathrm{this\, work}$ &  $E^\mathrm{GVU}$ & $A^\mathrm{this\, work}_\mathrm{a}$ &$A^\mathrm{GVU}_\mathrm{a}$ & $A^\mathrm{this\ work}_\mathrm{r}$ &  $A^\mathrm{GVU}_\mathrm{r}$  \\
        &       &      eV       &      eV   &  $10^{13}$~s$^{-1}$    & $10^{13}$~s$^{-1}$ & $10^{11}$~s$^{-1}$ & $10^{11}$~s$^{-1}$  \\
\hline
2s2p	&	0	&	359.0493	&	359.04	&	1.40[+00]	&	1.31[+00]	&	7.06	&	7.1	    \\
	    &	1	&	359.0649	&	359.05	&	1.38[+00]	&	1.31[+00]	&	7.06	&	7.11	\\
	    &	2	&	359.0992	&	359.09	&	1.35[+00]	&	1.31[+00]	&	7.06	&	7.11	\\
2s3p	&	0	&	418.7346	&	418.74	&	5.92[-01]	&	6.12[-01]	&	4.74	&	4.64	\\
	    &	1	&	418.7455	&	418.75	&	5.84[-01]	&	6.17[-01]	&	4.65	&	4.51	\\
	    &	2	&	418.7658	&	418.77	&	5.80[-01]	&	6.23[-01]	&	4.47	&	4.26	\\
2p3s	&	0	&	419.1274	&	419.12	&	2.33[-02]	&	1.83[-02]	&	6.19	&	5.65	\\
	    &	1	&	419.1373	&	419.13	&	1.97[-02]	&	1.43[-02]	&	6.29	&	5.66	\\
	    &	2	&	419.1602	&	419.15	&	1.91[-02]	&	7.44[-03]	&	6.49	&	6.02	\\
2p3d	&	0	&	423.3094	&	423.36	&	3.79[-05]	&	1.86[-03]	&	6.69	&	6.47	\\
	    &	1	&	423.3029	&	423.35	&	3.35[-04]	&	2.20[-03]	&	6.68	&	6.82	\\
	    &	2	&	423.2922	&	423.34	&	5.48[-05]	&	1.91[-03]	&	6.67	&	6.75	\\
\end{tabular}
\end{ruledtabular}

\end{table*}

In Table~\ref{tab:comptheories} the MBPT results for excitation energies, Auger rates and rates for radiative decay to $1s n\ell$ states from $2s2p$, $2s3p$, $2p3s$, and $2p3d$ levels are compared  to data obtained by Goryaev \textit{et al.}~\cite{Goryaev2017} applying the $Z$-expansion method. Previous results~\cite{Vainshtein1978,Vainshtein1980} were superseded by the new improved calculations which, among other refinements, account for relativistic corrections within the framework of the Breit operator. The comparison shows that the excitation energies from the present theory and the calculations of Goryaev \textit{et al.} agree within deviations of at most 0.05~eV, i.e., within 120 ppm. In this context, one has to keep in mind that the uncertainties of the present, calculated resonance energies are below 1~meV, i.e., definitely less than 3~ppm. The largest deviations are found for the $2p3d$ energies. For the other resonances the maximum deviation is 0.015~eV corresponding to 36 ppm. The comparison is similarly satisfying for the radiative rates with a maximum difference of 11\%. For the $2s2p$ and $2s3p$ resonances, deviations of autoionization rates are within 7\%. The deviations increase for the $2p3s$ resonances and reach a factor of almost 2.6 and for the $2p3d$ resonances there are differences in the two calculations of Auger decay rates by factors reaching almost up to 50. While the latter numbers could not be tested by the present experiments because of the too small signal rates, experimental results were obtained for $2s2p$, $2s3p$, and $2p3s$ resonances. The measurements support the present theoretical resonance strengths which are immediately related to the radiative and Auger decay rates.

\begin{figure}
\includegraphics[width=\columnwidth]{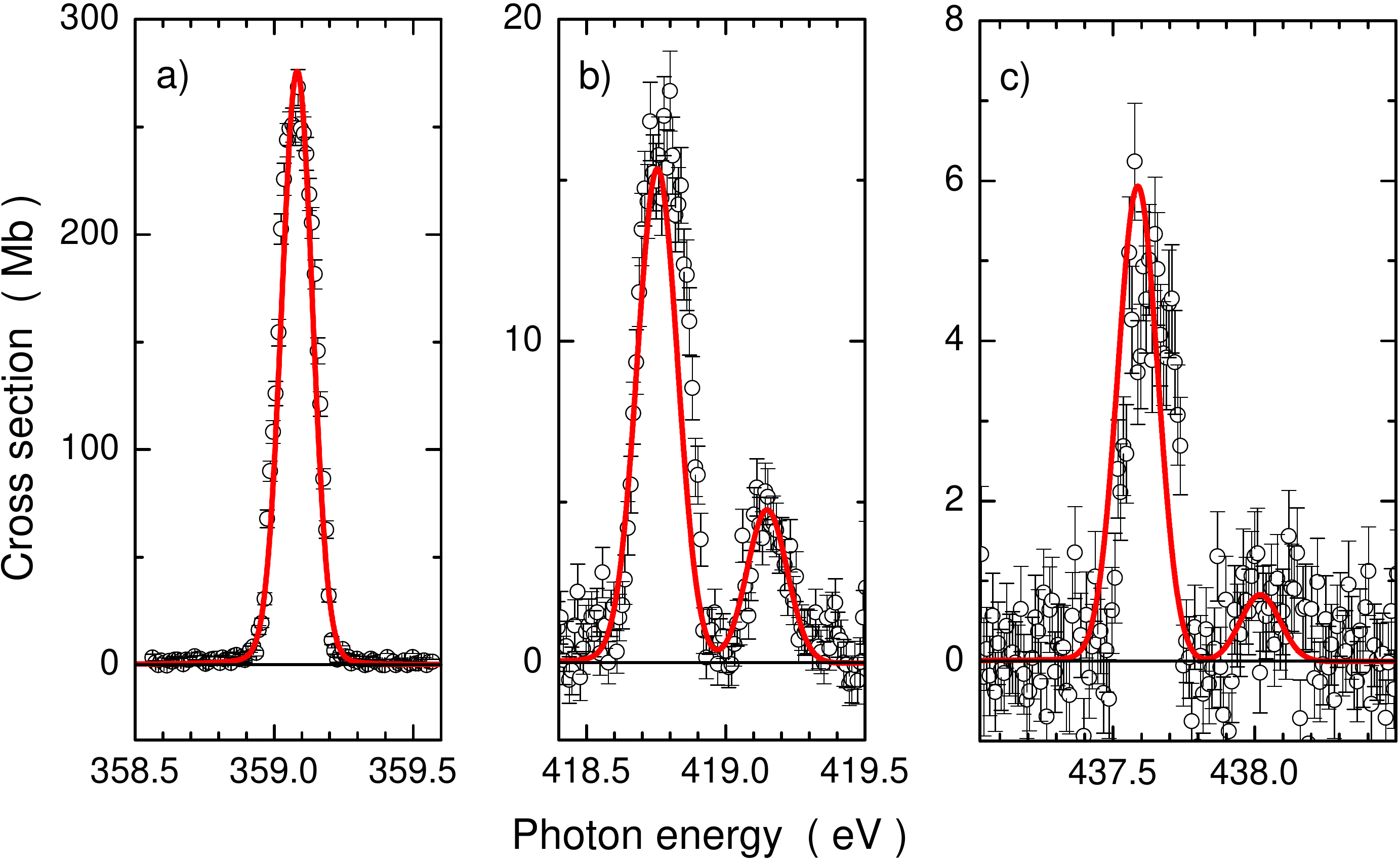}
\caption{\label{Fig:overview} (color online)  Cross sections for photoionization of C$^{4+}(1s2s~^3S)$ ions. The experimental data with statistical error bars were obtained at constant 500~{$\mu$}m exit-slit width of the monochromator. The solid red line is the present theory convoluted with Gaussians of appropriate FWHM values: a) 116~meV; b) 167~meV; c) 165~meV. The experimental spectrum was obtained in one multi-range scan and normalized such that the area of the first peak matches the theoretical resonance strength.
}
\end{figure}

By a number of experimental multi-range energy scans at a fixed monochromator exit-slit width of 500~$\mu$m, an overview of the resonance features within the energy region of interest was obtained at a resolving power of approximately 2800. Three groups of resonances in the energy ranges 358.6 to 359.6~eV, 418.4 to 419.5~eV, and 437.0 to 438.5~eV, respectively, were covered in a single scan. Such scans were repeated several times to improve the signal-counting statistics. For the $2p3d$ and $2p4d$ resonances the predicted resonance strengths for ionization  are about three orders of magnitude below the strengths of the $2s2p$ resonances (see Table~\ref{tab:resonanceparameters}). Due to low count rates and relatively high background, those small contributions could not be measured. Figure~\ref{Fig:overview}a shows the result of the combined overview scans covering the lower-energy region with the three $2s2p~^3P_{0,1,2}$ resonances. The photon energy scale was calibrated following the procedure described at the end of Sec.~\ref{Sec:experiment}.  The  experimental overview-scan spectrum was obtained on a relative scale and had initially only been nomalized to the ion current and the photon flux. The whole spectrum was then multiplied by a constant factor such that the area of the measured $2s2p$ peak shown in Fig.~\ref{Fig:overview}a matches the theoretical ionization strength. The solid red line in this figure was obtained by convoluting the theoretical spectrum with a Gaussian (the experimental response function at 500~$\mu$m monochromator exit-slit width does not appear to be strictly Gaussian, though). The set of three Gaussian-convoluted Fano profiles for the three $^3P$ resonances that resulted from the present theory was adjusted to the experimental data in a least-squares fit by leaving a global energy shift and the Gaussian width free for the fit while all other parameters were kept fixed at the values provided in Table~\ref{tab:resonanceparameters}. By this procedure the Gaussian width was determined to be 116~meV and the overall shift of the experimental to the theoretical $2s2p$ spectrum was found to be -1.4~meV.  In Fig.~\ref{Fig:overview} all data are shown without any energy shifts, i.e.,  the experimental and theoretical spectra are based on independent energy scales.

Figure~\ref{Fig:overview}b displays the experimental and theoretical photoionization cross section in the mid-energy region around 419~eV covering the  contributions of $2s3p~^3P_{0,1,2}$ and $2p3s~^3P_{0,1,2}$ resonances. As mentioned in the preceding paragraph, the experimental overview spectrum as a whole was normalized to theory at the 359-eV peak by an energy-independent factor. A fit of the experimental data similar to the one applied to the peak at 359~eV  suggests a bandwidth of 167~meV and a global shift of the $2s3p~^3P_{0,1,2}$ and $2p3s~^3P_{0,1,2}$ resonance group relative to the present theory by +19.0~meV. As in  Fig.~\ref{Fig:overview}a, the theoretical data, convoluted with a 167-meV FWHM Gaussian, and the experimental cross sections are shown on their individual, intrinsically determined energy scales. It is worth noting that the ``apparent'' cross sections (after convolution with the experimental response function) are down in peak height from the (also convoluted) $2s2p$ resonances  by approximately a factor of 15.

The data displayed in Fig.~\ref{Fig:overview}c are yet another factor of about 3 down from those in Fig.~\ref{Fig:overview}b. The statistical scatter is relatively large. Nevertheless, signal from the $2s4p~^3P_{0,1,2}$ and $2p4s~^3P_{0,1,2}$ resonances could be observed. Again, as in Fig.~\ref{Fig:overview}b, the experimental spectrum is shown on the experimentally determined energy scale without any further manipulation beyond the normalization to the $2s2p$ peak at 359~eV. A fit of the experimental data similar to the one  applied to the peak at 359~eV  suggests a bandwidth of 165~meV and a global shift of the $2s4p~^3P_{0,1,2}$ and $2p4s~^3P_{0,1,2}$ resonance group relative to the present theory by +40.5~meV. Accordingly, the present theoretical results were convoluted with a 165-meV FWHM Gaussian for the comparison with the experimental data. The differences of the derived photon-energy spreads are observed at a constant monochromator exit-slit width of 500~$\mu$m. They are in the range of a resolving power of 2950$\pm$450. Excursions from a constant resolving power can  partly be attributed to statistical fluctuations and partly to the fact that the photon optics were optimized for the resonance group at 359~eV and the quality of the settings drifted away from the optimum with increasing photon energy.

The theoretical and experimental results in Fig.~\ref{Fig:overview} are remarkably consistent. Only a single constant factor was applied to the experimental scan spectrum for normalization to the strength of the $2s2p$ ionization peak and the maximum deviation between experimental and theoretical resonance energies is 40.5~meV at about 438~eV. As discussed in some detail at the end of Sec.~\ref{Sec:experiment} all the deviations of experimental and theoretical resonance positions are well within the estimated uncertainties of the experimental energy calibration.

\begin{figure}
\includegraphics[width=\columnwidth]{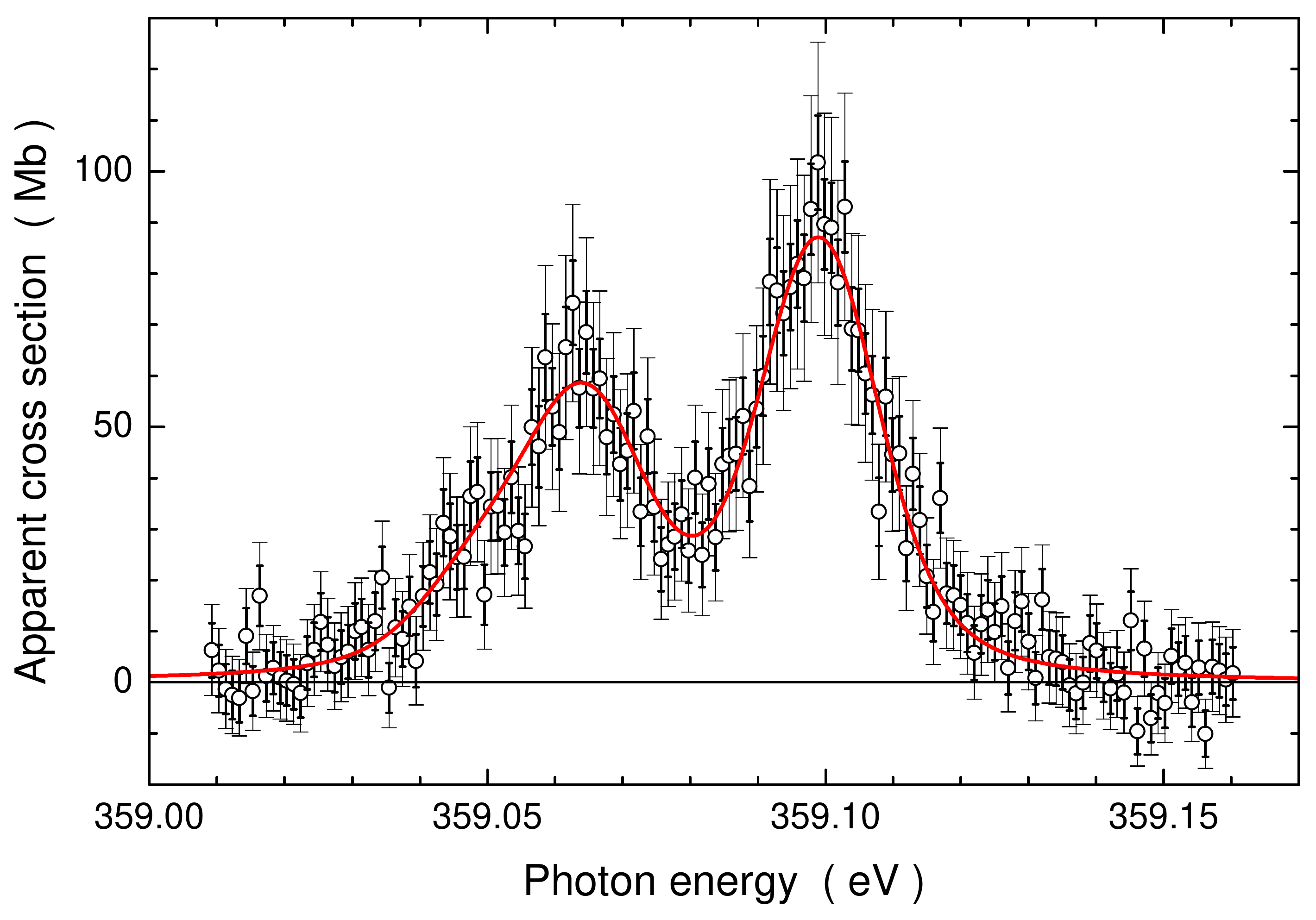}
\caption{\label{Fig:absolutescan} (color online)  Apparent absolute cross sections for photoionization of C$^{4+}(1s2s~^3S)$ ions via $2s2p~^3P_{0,1,2}$ resonances. The experimental energies were shifted up by 1.4~meV to match the theoretical resonance positions. Each experimental data point is shown with its statistical and  its combined statistical and systematic, i.e. total uncertainty. The measurement was carried out with a C$^{4+}$ ion beam containing an unknown fraction $f$ of $^3S$ metastable ions. The solid red line represents the present theoretical cross section after convolution with a 16.2-meV FWHM Gaussian and multiplication by a factor 0.105 which is interpreted to represent~$f$.
}
\end{figure}

The experimental setup is well suited for absolute cross section measurements and the energy resolution can be substantially enhanced by closing the exit slit of the monochromator (at the expense of photon flux). The experimental conditions for absolute measurements of cross sections are described in Sec.~\ref{Sec:experiment}. The result of the absolute measurement of the apparent cross section for the $2s2p~^3P_{0,1,2}$ resonances at a photon-energy bandwidth of 16.2~meV (20~$\mu$m exit-slit width) is displayed in Fig.~\ref{Fig:absolutescan}. Both the statistical and absolute error bars of each point are shown. Obviously, the $2s2p~^3P_{2}$ resonance at 359.099~eV has been clearly resolved from the peak consisting of the two $2s2p~^3P_{0}$ and $^3P_{1}$ resonances. The solid red line is the present theoretical result convoluted with a 16.2-meV FWHM Gaussian and multiplied with a factor of 0.105. At this point, one has to recall that the C$^{4+}$ ion beam used in the experiment contained unknown fractions $f$ of metastable C$^{4+}(1s2s~^3S)$ and $1-f$ of ground-state C$^{4+}(1s^2~^2S)$ ions. The peak structure displayed in Fig.~\ref{Fig:absolutescan} is due to $2s2p~^3P$ resonances which are exclusively populated by excitation starting from the metastable beam fraction. Under the assumption, that the present theory is correct, the apparent ionization cross section found in the experiment is $f$ times that obtained by theory. Thus, the factor 0.105 found in the comparison of theory and experiment has to be interpreted as the fraction $f$ of metastable ions in the parent beam. The statistical scatter of the individual data points is about 10\% at the cross-section maximum. However, the statistical uncertainty of the total strength contained in the $2s2p~^3P$ resonances is only 2.6\%. Thus the fraction $f$ is determined with a statistical uncertainty of only 0.003 out of 0.105, i.e., $f = 0.105 \pm 0.003$. The absolute uncertainty is about 15\% due to the systematic uncertainties of the cross-section measurements.

\begin{figure}
\includegraphics[width=\columnwidth]{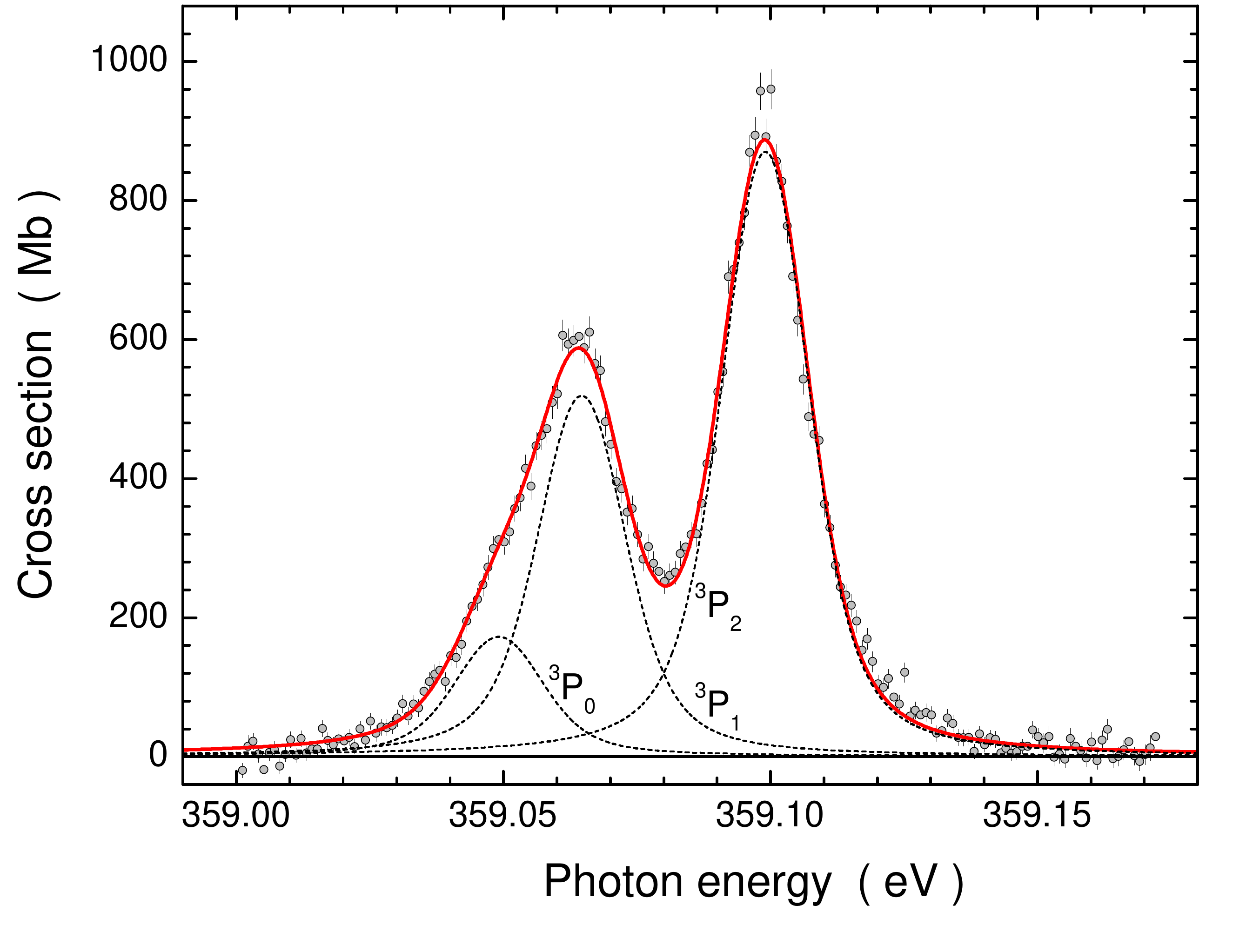}
\caption{\label{Fig:scan2s2p} (color online) High-resolution cross section for photoionization of C$^{4+}(1s2s~^3S)$ ions via $2s2p~^3P_{0,1,2}$ resonances. The experimental cross sections with statistical error bars are normalized to 100\% $^3S$ metastable parent ions. The solid red line represents the present theoretical cross section after convolution with a 14.38-meV FWHM Gaussian which corresponds to a resolving power $E/\Delta E = 25,000$. The individual contributions of the fine-structure components are shown by thin dotted lines. The experimental spectrum was shifted up in energy by 1.4~meV.
}
\end{figure}

In an absolute measurement only a defined part (about 50~cm)  of the total photon-ion interaction length (approximately 175~cm between the merger and the demerger) is used. Hence, the count rates in an absolute measurement are reduced, typically by a factor of 2,  and it takes more time to accumulate counts for a desired level of statistics. A more economical way to get better statistics is to use the whole interaction path length and, by that, obtain a relative cross section which can then be normalized to a measurement like the one displayed in Fig.~\ref{Fig:absolutescan}. For best possible energy resolution,  the fixed-focus constant c$_\mathrm{ff}$~\cite{Viefhaus2013,Reininger2005} of the variable-line-spacing (VLS) grating in use at beamline P04 was carefully adjusted, and thus the bandwidth could be reduced to 14.38$\pm$0.23~meV. Under these conditions, 19 sweeps over the energy range 359.00 to 359.17~eV with 1~meV step size and 10~s dwell time each were accumulated during several days. The resulting relative spectrum was normalized to the absolute measurement shown in Fig.~\ref{Fig:absolutescan} and then normalized again to 100\% metastable C$^{4+}(1s2s~^3S)$  ions in the parent ion beam, i.e., the normalized spectrum was divided by $f = 0.105$. The statistical uncertainties of each data point at the cross section maximum are now down to less than 3\%.  Figure~\ref{Fig:scan2s2p} shows the final experimental data and, as a solid red line,  the result of the present cross-section calculation convoluted with a 14.38-meV FWHM Gaussian. At the resulting resolving power of 25,000 the presence of the $2s2p~^3P_{0}$ contribution starts to show as a hump on the low-energy side of the $^3P_1$ peak. No  photoionization experiment with ions resolving fine-structure components of a given deep-inner-shell resonance term at such high energies has been reported so far~\cite{Mueller2015c}.

\begin{figure}
\includegraphics[width=\columnwidth]{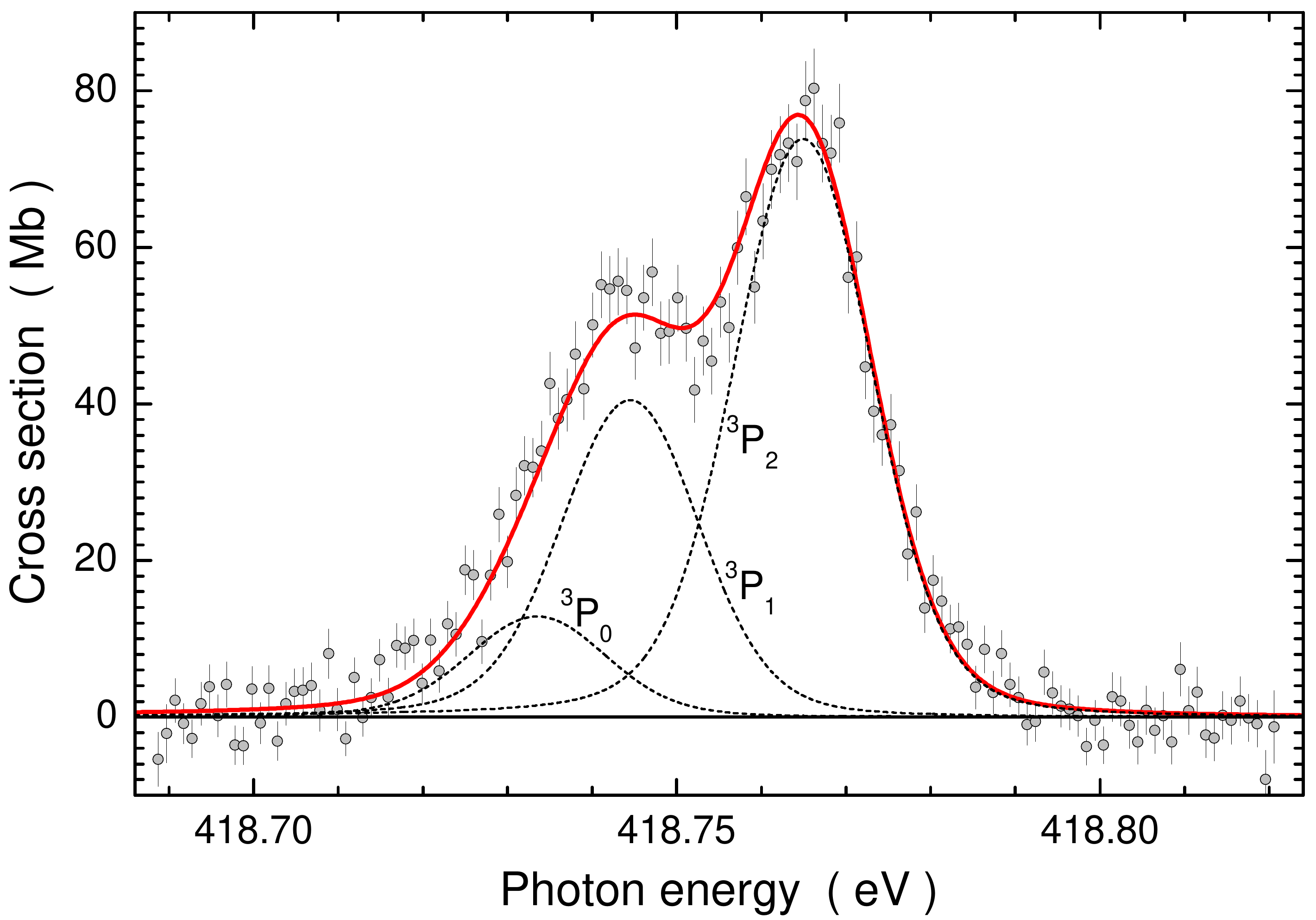}
\caption{\label{Fig:scan2s3p} (color online) Cross sections for photoionization of C$^{4+}(1s2s~^3S)$ ions via $2s3p~^3P_{0,1,2}$ resonances. The experimental cross sections with statistical error bars are normalized to 100\% $^3S$ metastable parent ions. The solid red line represents the present theoretical cross section after convolution with a 17.17-meV FWHM Gaussian which corresponds to a resolving power $E/\Delta E = 24,400$ similar to the one in Fig.~\ref{Fig:scan2s2p}. The individual contributions of the fine-structure components are shown by thin dotted lines. The experimental spectrum was shifted down in energy by 19~meV.
}
\end{figure}

Similar energy-scan measurements were performed for the $2s3p~^3P_{0,1,2}$ resonances in the energy range 418.689 to 418.821~eV at 1~meV step size and at comparable resolving power. The energy spread was determined by a fit yielding 17.17$\pm$0.50~meV. The experimental peak area was normalized to the known ratio of the $2s3p$ and $2s2p$ photoionization resonance strengths (see Fig.~\ref{Fig:overview}) considering the absolute apparent strength of the $2s2p$ resonance (see Fig.~\ref{Fig:absolutescan}) and the fraction $f$ of metastable parent ions in the beam. The absolute cross section for (100\%) metastable C$^{4+}(1s2s~^3S)$ ions thus obtained is compared in Fig.~\ref{Fig:scan2s3p} with the theoretical data convoluted with a 17.17-meV FWHM Gaussian. As for the $2s2p$ resonances (see Fig.~\ref{Fig:scan2s2p}) theory and experiment are in excellent agreement with the assumption of a metastable fraction $f$ of 10.5\%. At the present resolving power the $2s2p~^3P_{2}$ resonance is partly resolved from the $2s2p~^3P_{0,1}$ sum peak.

The present results on photoionization of C$^{4+}(1s2s~^3S)$ can be tested against previous absolute measurements on dielectronic recombination (DR)  of C$^{5+}(1s)$ ions~\cite{Wolf1991}. The basis for the comparison is the principle of detailed balance~\cite{Flannery2006,Mueller2002b,Schippers2002b,Schippers2003a,Schippers2004b,Mueller2009a,Mueller2010a,Mueller2014a} which, in turn, is based on time-reversal symmetry in atomic processes. The present theoretical calculations of C$^{4+}$ photoionization cross sections also provide all the information to  directly infer  C$^{5+}$ DR cross-section contributions proceeding via resonances that are associated with C$^{4+}(n\ell n'\ell'~^3P_{0,1,2})$ doubly excited states with $n=2$ and $n'=2,3$,
\begin{eqnarray}
 \label{Eq:CVIDR}
e + C^{5+}(1s~^2S_{1/2}) \to ~ &   C^{4+}(n\ell n'\ell'~^3P_{0,1,2}) \nonumber \\
& \downarrow\\
& C^{4+}(1s2s~^3S_1) + \gamma .   \nonumber
\end{eqnarray}
The present calculations were extended to include all the ten possible C$^{4+}(2\ell 2\ell')$ product levels with $\ell, \ell' = s, p$ that can be populated in DR of C$^{5+}(1s~^2S)$ ions.

The C$^{5+}$ DR experiment~\cite{Wolf1991} was one of the first such measurements carried out at a heavy-ion storage ring with an electron cooling device. In the early stage of the DR measurements the energy resolution at energies around 270~eV was limited and amounted to approximately 2.5~eV. Later developments of cold electron targets~\cite{Wolf2009} have substantially improved the energy resolution in DR experiments. With  present state-of-the-art experimental equipment the energy spread in the electron-ion center-of-mass system could be as low as 0.25~eV at 270~eV with a longitudinal electron beam temperature $k_\mathrm{B}T_\parallel = 2 \times 10^{-5}$~eV~\cite{Wolf2009,Mueller1999a} where $k_\mathrm{B}$ is Boltzmann's constant.

The only existing DR experiment~\cite{Wolf1991} covered the electron energy range 260 to 380~eV which includes all resonances associated with $2\ell n\ell'$ configurations with $n=2,3,4,...; \ell, \ell' = 0,1,...,n-1$. At the limited resolution of the experiment, structure within these configurations could only be observed for $n=2$. All the ten possible levels within the $2\ell 2\ell'$ configurations contribute to DR of C$^{5+}(1s~^2S_{1/2})$. However, only the $2s2p~^3P$ term can contribute to photoionization of C$^{4+}(1s2s~^3S)$ due to the selection rules for electric dipole transitions. Hence, for a comparison of DR of C$^{5+}(1s~^2S_{1/2})$ with photoionization of C$^{4+}(1s2s~^3S_1)$ and for exploiting time-reversal symmetry, only the $2s2p~^3P_{0,1,2}$ resonances can be considered.

Time reversal symmetry and the principle of detailed balance relate the cross section $\sigma^{\mathrm{DR}}$ for the DR process characterized by Eq.~\ref{Eq:CVIDR} with the cross section $\sigma^{\mathrm{PI}}$ for the photoionization process described by Eq.~\ref{Eq:CVPI} with $n, n' = 2$ on a level-to-level basis,
\begin{equation}\label{eq:balance}
   \frac{ \sigma^{\mathrm{DR}}}{\sigma^{\mathrm{PI}}} =
     \frac{g_i}{g_f}\frac{E_\gamma^2}{2m_\mathrm{e} c^2 E_\mathrm{e}},
\end{equation}
where the quantities $g_i=g_i(1s2s~^3S)=3$ and $g_f=g_f(1s~^2S)=2$ are the statistical weights of the initial and final levels of the photoionization process,
respectively, and $m_\mathrm{e}$ is the electron rest mass. The photon energy $E_\gamma$ and the electron energy $E_\mathrm{e}$ in the  processes related to one another by time reversal are connected by the condition
\begin{equation}
E_\mathrm{e} = E_\gamma - I_{\mathrm{bind}},
\end{equation}
where $I_{\mathrm{bind}}=93.131$~eV is the ionization energy of the C$^{4+}(1s2s~^3S_1)$ level~\cite{NIST2018}. Thus, the $2s2p~^3P$ resonance group found at about 359~eV in photoionization of C$^{4+}(1s2s~^3S_1)$ has to appear at about 266~eV in DR of C$^{5+}(1s~^2S_{1/2})$.

The DR cross section for a a transition $i \to j \to f$ from an initial state $i$ to a final state $f$ via an intermediate resonance $j$ can be written as~\cite{Griffin1989b}
\begin{eqnarray}
\label{eq:DRWQ}
\sigma^{\mathrm{DR}} = \frac{\hbar^3}{m_\mathrm{e}}\frac{\pi^2}{2 E_\mathrm{e}}\frac{g_j}{g_i}   \frac{A_\mathrm{a}(j\to i) \sum_{f'}{A_\mathrm{r}(j \to f')}}{\Gamma(j)/\hbar} \times \\      \nonumber
  \frac{1}{2\pi}\frac{\Gamma(j)}{(E_\mathrm{e}-E_\mathrm{res})^2+\Gamma(j)^2/4}
\end{eqnarray}
where $g_j$ is the statistical weight of the resonant level $j$, $A_\mathrm{a}(j\to i)$ is the Auger decay rate of the resonant level $j$ to the initial level $i$, $A_\mathrm{r}(j \to f')$ comprises all radiative rates for the resonant level $j$ decaying to all bound levels $f'$, $E_\mathrm{res}$ the resonance energy at which level $j$ is populated, and the total natural width of the resonance $j$,
\begin{equation}
\Gamma(j) = \hbar\left[\sum_{i'}{A_\mathrm{a}(j \to i')}+\sum_{f'}{A_\mathrm{r}(j \to f')}\right].
\end{equation}

\begin{figure}
\includegraphics[width=\columnwidth]{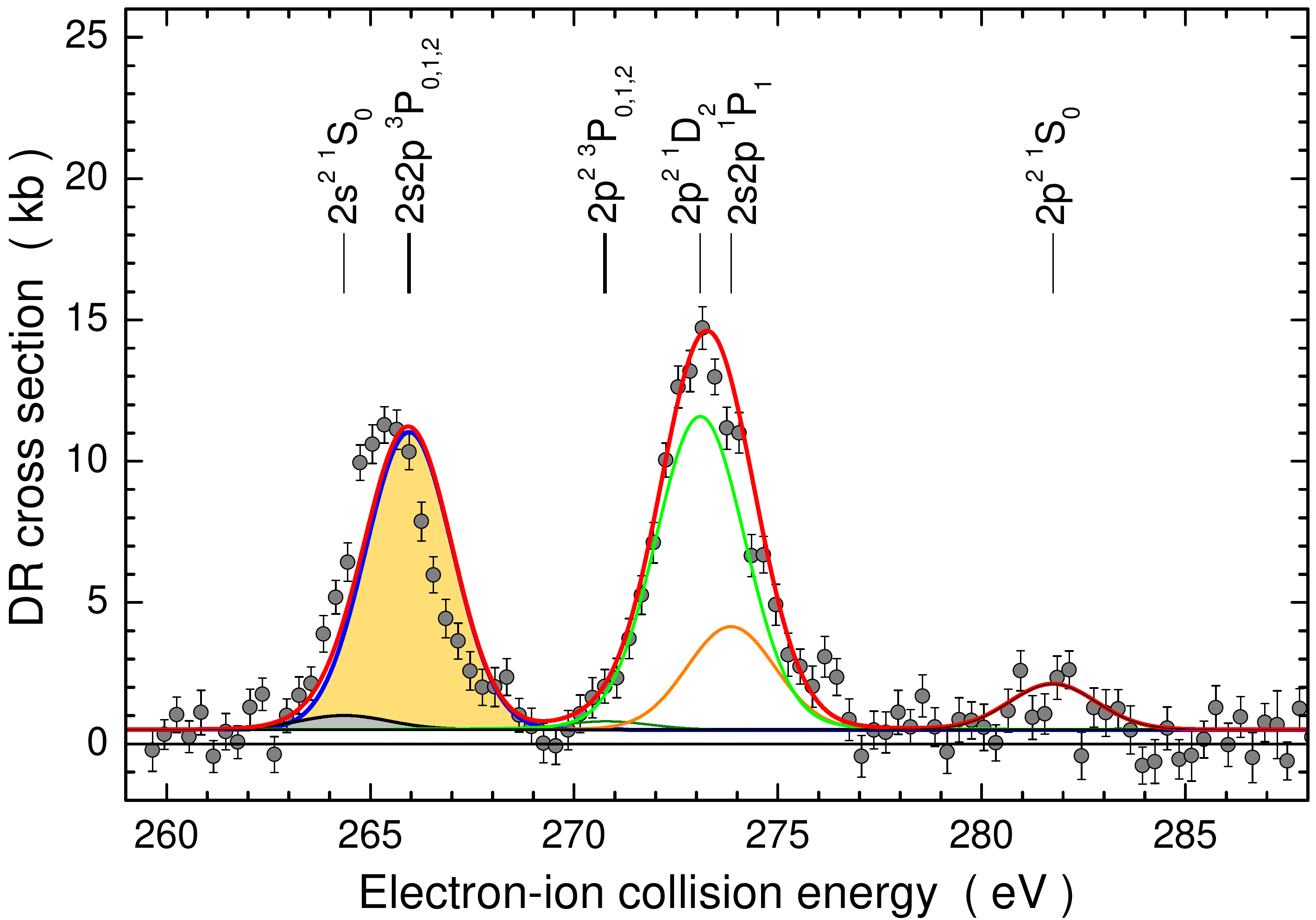}
\caption{\label{Fig:CVIDR} (color online)  Cross sections for dielectronic recombination (DR) of C$^{5+}(1s~^2S_{1/2})$ in the energy range of $2\ell 2\ell'$ resonances. The solid circles with statistical error bars show the results obtained at a heavy-ion storage ring~\cite{Wolf1991}. The solid red line represents the present DR calculation. The theoretical cross-section function has been convoluted with a 2.5-eV FWHM Gaussian to simulate the experimental energy spread.  The energies of the individual resonance contributions in the present energy range are shown by vertical bars and the associated doubly excited levels are indicated.
The individual-term contributions to the theoretical DR spectrum are shown by Voigt profiles which can be identified by their resonance energies. In particular, the solid black line shows the contribution of the $2s^2~^1S_0$ resonance and the blue line represents the contribution of the $2s2p~^3P^o$ term. For comparison, the contributions of the $2s^2~^1S$ resonance (gray shaded peak) and the $2s2p~^3P^o$ term (gold shaded peak)  were inferred from the decay rates provided by Goryaev \textit{et al.}~\cite{Goryaev2017}. On the scale of the figure, the latter contributions cannot be distinguished from the present theoretical results.
}
\end{figure}

\begin{table*}
\caption{\label{tab:detailedbalance}
DR  strengths in kb\,eV of the $2s2p~^3P^o$ and $2s^2~^1S$ resonance terms in $^{12}$C$^{4+}$ and their sum $\Sigma$ determined in different experiments and by theoretical calculations. For the fraction $f$ of C$^{4+}(1s2s~^3S)$ metastable ions in the parent ion beam used in the present photoionization experiments the comparison with the present photoionization theory suggests $f=0.105$. This fraction has been used for calculating the entry of $S(2s2p~^3P) = 28.1$~kb\,eV in the last row.  Comparison with the DR measurement instead would have resulted in $f=0.108$.
}
\begin{ruledtabular}
\begin{tabular}{llll}
data source & $S(2s2p~^3P)$ & $S(2s^2~^1S)$ & $\Sigma$\\
\hline
Present MBPT and CR theory   & 28.1 & 1.4 & 29.5 \\
Inferred from Z-expansion decay rates~\cite{Goryaev2017} & 28.3  & 1.4  & 29.7  \\
DR experiment with C$^{5+}$ ions~\cite{Wolf1991} & 28.7-1.4=27.3 & & 28.7 \\
present time-reversed photoionization  & 28.1  & & \\
\end{tabular}
\end{ruledtabular}
\end{table*}

 Figure~\ref{Fig:CVIDR} shows the DR results~\cite{Wolf1991} obtained in the range of the $2\ell 2\ell'$ resonances. The  data have been corrected for an error in the original analysis (see~\cite{Badnell2006a}). The ten possible levels within the $2\ell 2\ell'$ manifold are partly resolved. Three peaks can be identified. The energies and Auger rates of the $2\ell 2\ell'$ resonances have been calculated within the framework of the present theory and the radiative decay rates have been calculated within first order perturbation theory as previously discussed. All the quantities needed for Eq.~\ref{eq:DRWQ}  are thus at hand.  The result is displayed  in Fig.~\ref{Fig:CVIDR}, where the energy positions are indicated by the vertical bars  and labeled by the spectroscopic notation of the associated levels. A  detailed account of the present method for calculating DR cross sections has been provided previously by Tokman \textit{et al.}~\cite{Tokman2002}.  The individual cross-section contributions of all levels have also been calculated. Solid lines in Fig.~\ref{Fig:CVIDR} represent the contributions of the six terms associated with $2\ell 2\ell'$ configurations. The sum of these contributions (the red line) provides a very good representation of the experimental peak areas. Deviations of experimental and theoretical resonance energies illustrate the limitations of the early DR measurement for which an energy uncertainty of $\pm1$~eV at 500~eV was quoted.

 In addition to the present theoretical calculations and for comparison, the DR contributions arising from the $2s^2~^1S$ and $2s2p~^3P^o$ resonance terms were inferred from the decay rates provided by Goryaev \textit{et al.}~\cite{Goryaev2017} using Eq.~\ref{eq:DRWQ}. Since the DR spectrum was  measured at an electron-energy bandwidth of approximately 2.5~eV the cross section obtained via Eq.~\ref{eq:balance} has to be convoluted with a 2.5-eV FWHM Gaussian to simulate the experimental conditions of the DR measurement. Excellent agreement of the resonance energies and resonance strengths resulting from the two independent theoretical approaches is observed.

 Table~\ref{tab:detailedbalance} provides DR resonance strengths $S$ in kb\,eV for DR of C$^{5+}(1s^2S_{1/2})$ leading to the $2s2p~^3P^o$ and $2s^2~^1S$ resonance terms together with the sum $\Sigma = S(2s2p~^3P) + S(2s^2~^1S)$ determined by different experiments and theoretical calculations. It is obvious from Fig.~\ref{Fig:CVIDR} that the first peak in the spectrum is a blend of contributions from the $2s^2~^1S$ and $2s2p~^3P^o$ terms. The present theoretical calculations and the strengths inferred from the decay rates determined by Goryaev \textit{et al.}~\cite{Goryaev2017} provide a very consistent picture with almost identical results. Both theoretical approaches predict a strength $S(2s^2~^1S) = 1.4$~kb\,eV which is only about 5\% of the strength $S(2s2p~^3P) \approx 28$~kb\,eV.

 The DR experiment with C$^{5+}$ ions~\cite{Wolf1991} yields $S(2s2p~^3P) + S(2s^2~^1S) = 28.7$~kb\,eV. The uncertainty of this strength is directly related to the systematic uncertainty of the experimental cross sections which has been quoted to be $\pm15$\%. By subtraction of the theoretically predicted, relatively small contribution $S(2s^2~^1S) = 1.4$~kb\,eV the strength  $S(2s2p~^3P) = 27.3$~kb\,eV results. This number has to be compared with the strength  $S(2s2p~^3P) = 28.1$~kb\,eV inferred from the present photoionization experiments and by exploiting the principle of detailed balance. The difference is less than 3\% and thus very much smaller than the systematic uncertainties of both the present photoionization and the previous DR cross-section measurements.

The comparison of the present photoionization experiments with the independent measurement of DR cross sections provides additional support for the present photoionization data and, hence, also for the fraction $f=0.105$ of $1s2s~^3S$ metastable ions in the C$^{4+}$ beam that was employed in the photoionization experiments.

\section{Summary and outlook}
In the present photoionization experiments with metastable C$^{4+}(1s2s~^3S_1)$ ions the energy range of doubly excited (empty-\textit{K}-shell) $2s2p, 2s3p, 2p3s, 2p3d, 2s4p, 2p4s$, and $2p4d$ resonances was investigated and absolute apparent cross sections were measured with a mixed beam of ground-state and metastable $^3S$ ions. By comparison with the results of the present relativistic many-body-perturbation theory (RMBPT) it was possible to determine the fraction $f = 10.5$\% of the $^3S$ metastable component. The resolving power (25,000) of the experiment was sufficient to resolve  fine-structure in the doubly excited $2s2p~^3P_{0,1,2}$ and $2s3p~^3P_{0,1,2}$ resonance manifolds. The experimental resonance energies agree with the present RMBPT calculations  within the experimental uncertainties. The maximum deviation of experiment from theory is about 40~meV at 437~eV. Given the fact that the fraction $f$ of the metastable component of the ion beam was not independently determined, the present experiment could only test the relative sizes of the dominant theoretical resonance contributions. Apart from a constant factor that is associated with the metastable fraction $f$ theoretical and experimental photoionization cross sections are in excellent agreement. A test of the theoretical C$^{4+}(2s2p~^3P)$ photoionization resonance contribution on an absolute scale is possible by comparison with measured absolute cross sections for dielectronic recombination of C$^{5+}$ and exploiting the principle of detailed-balance for time-reversed processes. This comparison shows that a fraction $f=0.105$ is consistent with both the C$^{4+}$ photoionization and the previous,  independent  C$^{5+}$ dielectronic recombination experiments.

A particular effort was made to determine the resonance energies with high accuracy. For this purpose, the RMBPT calculations were performed including all orders. Relativistic effects as well as QED up to the level of second order contributions were included. The energies thus obtained are estimated to have a maximum uncertainty of 0.001~eV. Such an accuracy makes the heliumlike C$^{4+}$ ion a promising candidate for being used as a primary reference standard for the soft-x-ray region with an uncertainty that is roughly a factor of one hundred better than the present neutral-gas standards in the energy range 300 to 1000 eV. It will be interesting to study ionization of heliumlike ions in the $1s2s~^3S_1$ level  along the associated very fundamental isoelectronic sequence. Further work on this topic is underway.

The very high accuracy of resonance energies obtained by the present theoretical treatment  may be exploited in an envisaged effort at the PIPE setup to generate new secondary reference standards for the calibration of soft-x-ray beamlines at synchrotron-radiation sources. A viable scenario would be the transfer of the calibration obtained from the very accurately calculated photoionization-resonance energies  of selected heliumlike (and lithiumlike~\cite{Yerokhin2017a,Yerokhin2017b}) ions to the neutral-gases-based calibration standards that are presently used. Full advantage of the existing high-precision theoretical resonance energies can only be taken if an  improved control of the photon-beam energy at beamline P04 can be realized which includes the stable positioning of the electrons circulating in the PETRA III storage ring.\\

\section{ACKNOWLEDGEMENTS}
This research was carried out in part at the light source PETRA III at DESY, a member of the Helmholtz Association (HGF). Support from Bundesministerium f\"{u}r Bildung und Forschung provided within the "Verbundforschung" funding scheme (contract numbers 05K10RG1,  05K10GUB, 05K16RG1, 05K16GUC) and from Deutsche Forschungsgemeinschaft under project numbers Mu 1068/22, Schi 378/12, and SFB925/A3  is gratefully acknowledged. S.K. acknowledges
support from the European Cluster of Advanced Laser Light Sources (EUCALL) project which has received funding from the European Union’s Horizon 2020 Research and Innovation Programme under Grant Agreement No. 654220. S.B. is supported by the Helmholtz Initiative and Networking Fund through the Young Investigators Program and by the Deutsche Forschungsgemeinschaft, project B03/SFB755. P.-M.H. acknowledges support by the Helmholtz-CAS Joint Research Group HCJRG-108. We  thank  F. Scholz and J. Seltmann for assistance in using beamline P04. Laboratoire Kastler Brossel (LKB) is ``Unit\'e Mixte de Recherche de Sorbonne Universit\'e, de ENS-PSL Research University, du Collège de France et du CNRS n$^{\circ}$ 8552''.
P.I. is a member of the Allianz Program of the Helmholtz Association, contract n$^{\circ}$ EMMI HA-216 ``Extremes of Density and Temperature: Cosmic Matter in the Laboratory''. E.L. acknowledges support from the Swedish Research Council, Grant No. 2016-03789.


%

\end{document}